\documentclass[aps,prc,twocolumn,groupedaddress]{revtex4-2}

\usepackage{graphicx}
\usepackage{braket}
\usepackage{amsmath}
\usepackage[dvipsnames]{xcolor}
\usepackage[normalem]{ulem} 

\begin{document}


\title{Two-neutrino double-beta decay matrix elements based on relativistic nuclear energy density functional}


\author{N. Popara}
\email[]{nato.popara@vef.unizg.hr}
\affiliation{Faculty  of  Veterinary Medicine,  University  of  Zagreb,  Heinzelova  c.  102,  10000  Zagreb,  Croatia}
\affiliation{Department  of  Physics,  Faculty  of  Science,  University  of  Zagreb,  Bijeni\v{c}ka  c.  32,  10000  Zagreb,  Croatia}

\author{A. Ravli\'c}
\email[]{aravlic@phy.hr}
\affiliation{Department  of  Physics,  Faculty  of  Science,  University  of  Zagreb,  Bijeni\v{c}ka  c.  32,  10000  Zagreb,  Croatia}

\author{N. Paar}
\email[]{npaar@phy.hr}
\affiliation{Department  of  Physics,  Faculty  of  Science,  University  of  Zagreb,  Bijeni\v{c}ka  c.  32,  10000  Zagreb,  Croatia}


\date{\today}

\begin{abstract}

Nuclear matrix elements (NMEs) for two-neutrino double-beta decay ($2\nu\beta\beta$) are studied in the framework of relativistic nuclear energy density functional (REDF). The properties of nuclei involved in the decay are obtained using the relativistic Hartree-Bardeen-Cooper-Schrieffer theory and relevant nuclear transitions are described using the relativistic proton-neutron quasiparticle random phase approximation based on relativistic energy density functional (REDF-QRPA). Three effective interactions have been employed, including density-dependent meson-exchange (DD-ME2) and point coupling interactions (DD-PC1 and DD-PCX), and pairing correlations are described consistently both in $T=1$ and $T=0$ channels using a separable pairing interaction. The optimal values of $T=0$ pairing strength parameter $V_{0pp}$ are constrained by the experimental data on $\beta$-decay half lives. The $2\nu\beta\beta$ matrix elements and half-lives are calculated for several nuclides experimentally known to undergo this kind of decay: $^{48}$Ca, $^{76}$Ge, $^{82}$Se, $^{96}$Zr, $^{100}$Mo, $^{116}$Cd, $^{124}$Xe, $^{128}$Te, $^{130}$Te, $^{136}$Xe and $^{150}$Nd. 
The model dependence of the NMEs and their sensitivity on $V_{0pp}$ is investigated, and the NMEs obtained using optimal values of $V_{0pp}$ are discussed in comparison to previous studies.
The results of the present work represent an important benchmark for the future applications of the relativistic framework in studies of neutrinoless double-beta decay.
\end{abstract}


\maketitle

\section{Introduction \label{Intro}}

The study of double-beta decays has attracted considerable interest over the past years. \cite{Pirinen2015,Engel2017}. Most of this interest has been focused on the neutrinoless mode ($0\nu\beta\beta$), because it offers the possibility of distinguishing between Dirac and Majorana nature of neutrinos \cite{Jiao2018}. The two-neutrino decay mode ($2\nu\beta\beta$), nonetheless, remains of interest, as an important benchmark of theoretical models for further studies of ($0\nu\beta\beta$) decays. The $2\nu\beta\beta$ decay is a second-order weak interaction process, and as such it is allowed by the Standard Model \cite{Haxton1984}, unlike the neutrinoless mode, which violates lepton conservation and consequently requires physics beyond the standard model \cite{Bilenky2015}. Furthermore, it is possible to experimentally observe this mode, as pairing in even-even nuclei makes them more stable than adjacent odd-odd nuclei, and the transition from an even-even mother to an odd-odd daughter is forbidden energetically, leaving double-beta decay as the only allowed decay channel \cite{Faessler1998}. This allows us to compare the results of our calculations to experiment directly, and, since the calculations for the neutrinoless mode use the same ingredients as those for the two-neutrino mode, their results can be used to constrain some of the model parameters and benchmark the model for studies of $0\nu\beta\beta$ decay \cite{Mustonen2013}.

The contribution of nuclear physics to the calculation of decay rates for both modes of double-beta decay is contained chiefly in the nuclear matrix element (NME) \cite{Staudt1990}. A variety of theoretical approaches has been developed in the past years for the description of double-beta decay NMEs. More detailed information is available in the following extensive reviews \cite{Primakoff1959,Tomoda1991,
Faessler1998,
Suhonen1998,Elliot2002,Avignone2008,Vergados2012,
Saakyan2013,Vergados2016,Engel2017,Dolinski2019,Ejiri2019}. 
The calculations of NMEs have been carried out using various approximations: the quasiparticle random phase approximation (QRPA) \cite{Suhonen2012,Terasaki2019,Ejiri20192}, its extension to the renormalized QRPA (RQRPA) \cite{Simkovic19972,Simkovic2009,Rodin2006,Stoica2001}, the second QRPA (SQRPA) \cite{Benes2006,Stoica2003}, various energy density functional (EDF) approaches \cite{Rodriguez2010,Menendez2014}, the Quasiparticle Tamm-Dancoff Approximation \cite{Ferreira2020}, the interacting shell model (ISM) and similar approaches \cite{Caurier2005,Kostensalo2020,Suhonen2019}, the interacting boson model (IBM) \cite{Barea2009,Brown2018,Barea2015},  and others \cite{Jiao2017,Rath2019,Kotila2010}. Among these approximations, the proton-neutron quasiparticle random-phase approximation (pn-QRPA) has emerged as one of the approaches that have been successfully employed in various studies of double-beta decays \cite{Simkovic1997,Pirinen2015,Unlu2018,Suhonen2011}. Within this approach, the usual isovector pairing in the ground state of open-shell nuclei is supplemented with an isoscalar proton-neutron pairing in the residual interaction. This has been shown to be important in nuclei relevant for double-beta decays (as first established by Ref. \cite{Vogel1986} and Ref.\cite{Engel1988} and confirmed by subsequent studies \cite{Martinez1999,Frauendorf2014}, including work beyond the QRPA approximation \cite{Hinohara2014,Menendez2016}).
In the relativistic framework, various methods have been employed in studies of $0\nu\beta\beta$ decays. In Ref. \cite{Song2014}  a
 multireference covariant EDF has been used to determine the wave functions of the initial and final nuclei, and correlations beyond the
 mean field have been described by configuration mixing of both angular momentum and particle number projected quadrupole
 deformed mean-field wave functions. Systematic study of $0\nu\beta\beta$ matrix elements in the relativistic EDF approach has been reported in Ref.~ \cite{Yao2015}, and effects of the the relativity and short-range correlations have recently been analyzed~\cite{Song2017}.
 
The aim of this work is to establish a theory framework for the description of $2\nu\beta\beta$ decay NMEs based on a modern relativistic nuclear energy density functional with density dependent point-coupling and meson-exchange interactions that include density dependence explicitly in the interaction vertex functions. These interactions have recently been introduced and successfully 
implemented in the description of nuclear excitation properties \cite{Paar2005,Paar2009,Niu2013,Niu2009,Khan2011,Yuksel2020}, astrophysically relevant weak interaction processes \cite{Samana2011,Fantina2012,Vale2016,Petkovic2019}, and nuclear equation of state \cite{RocaMaza2018,Mondal2016,Paar2014}. In previous relativistic EDF studies $2\nu\beta\beta$ decays
have not been addressed \cite{Song2014,Yao2015,Song2017}. We note that the present theory framework includes the density dependence 
explicitly in the vertex functions of the meson-exchange and point coupling interactions.

When modeling various quantities in nuclear physics, it is important to assess systematic errors. As pointed out in Ref.~\cite{Dobaczewski2014} the systematic error of a theoretical model may be a 
consequence of missing physics and/or poor modeling. Considering that in 
most cases the perfect model is not available, systematic errors are rather 
difficult to estimate. However, some insight about systematic 
uncertainties can be obtained from a comparative study of different theory 
frameworks and effective nuclear interactions. Therefore, it is important to
address the problem of double beta decays both from non-relativistic and relativistic 
frameworks, using various formulations of energy density functionals and different
parameterizations.

Since for the $2\nu\beta\beta$ decay mode experimental data exist for a set of nuclei, the present study also allows us to benchmark the relativistic model for the future 
studies of $0\nu\beta\beta$ decay.
The properties of nuclei involved are described using the relativistic Hartree-Bardeen-Cooper-Schrieffer (RH-BCS) model, while relevant transitions are obtained using the proton-neutron relativistic quasiparticle random phase approximation \cite{Paar2004}, that has recently been extended by including relativistic density-dependent point coupling interactions \cite{Vale2020,PhysRevC.104.064302}. In the following we denote this method as REDF-QRPA. Model calculations of the NMEs include
various double-beta emitters: $^{48}$Ca, $^{76}$Ge, $^{82}$Se, $^{96}$Zr, $^{100}$Mo, $^{116}$Cd, $^{124}$Xe, $^{128}$Te, $^{130}$Te, $^{136}$Xe, and $^{150}$Nd. 
In contrast to recent studies based on the pn-QRPA, which tend to start from a "realistic" nucleon-nucleon interaction~\cite{Fang2011,Hyvarinen2016}, in the present work in the particle-hole channel we use an interaction derived from a relativistic 
nuclear energy density functional. An important aspect of this study is that we are able to explore
the model dependence of the calculated NMEs, by implementing two different types
of relativistic density-dependent interactions, (i) finite range meson-exchange and (ii) point coupling interactions. Our model calculations also involve a treatment of the pairing correlations in open-shell nuclei, both at the level of the nuclear ground state and in the residual REDF-QRPA interaction. In particular, strength of the proton-neutron pairing, that is isoscalar $T=0$ pairing in the residual REDF-QRPA interaction has to be
constrained by using experimental data, in a similar way as already discussed in previous studies in non-relativistic frameworks \cite{Mustonen2013}. In the present study the experimental data on single $\beta$- decays will be employed to constrain $T=0$ pairing strength parameter 
for applications in $2\nu\beta\beta$ and in intended future $0\nu\beta\beta$ matrix element calculations.

We note that in this work, that represents the first implementation of the relativistic density dependent interactions in $2\nu\beta\beta$ decay study, some effects have not yet been taken into account, from deformation \cite{Delion2019,Fang2018} to renormalization and gauge symmetry \cite{Raduta2013} and isospin restoration \cite{Simkovic2018,Robledo2018}. In forthcoming studies more advanced effects will be taken into account. Nonetheless we expect that the present work will describe well at least some subset of the double-beta emitters, and will provide a useful guidance for the future studies. 

This article is organized as follows: in Section II we provide a theoretical overview concerning the matrix elements involved in double-beta decay and in Section III we outline the relativistic theory framework. Results are provided and discussed in Section IV, and the conclusion follows in Section V.

\section{Two-neutrino double-beta decay}

Two-neutrino double-beta decay ($2\nu\beta\beta$) is the process whereby two neutrons in the mother nucleus are converted into protons, accompanied by the emission of two electrons and two antineutrinos,

\begin{equation}
(A,Z) \rightarrow (A,Z+2) + 2e^- + 2\bar{\nu}.
\end{equation}

The half-life of two-neutrino double-beta decay formally depends on two matrix elements \cite{Simkovic2013},

\begin{equation}
\frac{1}{T_{1/2}^{2\nu}} = G^{2\nu}(Q,Z)g^4_A \left[\mathcal{M}^{2\nu}_{GT} + \mathcal{M}^{2\nu}_{F}\right]^2 ,
\label{halflife}
\end{equation}

where $\mathcal{M}^{2\nu}_{GT}$ and $\mathcal{M}^{2\nu}_{F}$ are, respectively, the Gamow-Teller (GT) and Fermi matrix element, and $G^{2\nu}(Q,Z)$ is a phase space factor that can be found tabulated, e.g., in Ref. \cite{Stoica2019}. The factor $g_A$ is the axial-vector coupling constant.

We restrict our consideration to decays from a 0$^+$ state to a 0$^+$ state. Decays to the states of higher angular momenta, e.g., 2$^+$, are possible, albeit highly suppressed. Their treatment would require a computational apparatus more involved than the REDF-QRPA used in this work \cite{Suhonen2007}. The Gamow-Teller matrix element in this case can be 
written as \cite{Ejiri2019}:

\begin{equation}
\label{eq2}
\mathcal{M}^{2\nu}_{GT} = m_e \sum_m \frac{ \bra{f}\sum_a \sigma_a \tau^-_a\ket{m}\bra{m}\sum_a \sigma_a \tau^-_a\ket{i} }{E_{m}+\frac{Q}{2}+m_e},
\end{equation}

where the sum goes over 1$^+$ states in the intermediate nucleus, labelled with $m$ and having the energy E$_m$ as measured from the ground state of the initial nucleus. 
The $Q$-value of the double-beta decay reads
\begin{equation}
Q = \left(m_i - m_f\right)c^2-2m_e.
\end{equation}
Note that we could have also chosen to group the factor $g_A^2$ into the definition of the matrix element in Eq. (\ref{eq2}), in order to make the discussion about the effects of $g_A$ and its renormalisation easier. If the two are grouped together, one obtains an effective NME ${M}^{2\nu}_{GT,eff}$ that depends on the choice of the axial-vector coupling as well \cite{Barabash20152}.
In this work, instead of the free nucleon value $g_A=1.26$, we employ a quenched value $g_A=1.0$ that is consistent with the previous REDF-QRPA studies of $\beta$-decay half-lives as well as electron capture rates \cite{Marketin2016,Ravlic2021, PhysRevC.102.065804}. Thus our results for the matrix elements in Eq. (\ref{eq2}) can be
 directly compared with those of the effective NMEs from Ref.~\cite{Barabash20152}.
Likewise, we introduce the factor m$_e$ (in MeV) in the definition of the nuclear matrix element to obtain dimensionless NMEs that can be compared easily to results of recent theoretical and experimental works. Otherwise the matrix elements would be in units of MeV$^{-1}$.
In all figures and tables in this work, the  REDF-QRPA results for ${M}^{2\nu}_{GT}$ and ${M}^{2\nu}_{GT,eff}$
are given as dimensionless.

The closure approximation is widely used in calculating $2\nu\beta\beta$ decay. It entails the replacement of the sum over different states in the intermediary nucleus, as in Eq. (\ref{eq2}), with a form that consists of one denominator containing a suitably-chosen "average" energy $\langle$E$\rangle$:

\begin{equation}
\mathcal{M}^{2\nu}_{GT, closure} = m_e \frac{ \bra{f}\sum_a \sigma_a \tau^-_a\sum_a \sigma_a \tau^-_a\ket{i} }{\langle E \rangle+\frac{Q}{2}+m_e}.
\label{eqclosure}
\end{equation}
In this work we calculate the NMEs using the more complete approach given in Eq. (\ref{eq2}), but where relevant for the comparison with other studies we include the closure approximation as well.
It can be shown that the Fermi matrix element vanishes if the same pairing interaction is consistently used at both the ground state and excitation levels in the isovector channel~\cite{Simkovic20182}. This is the case for our calculation and as such we will not show the Fermi matrix element.
The REDF-QRPA is not applicable to transitions from the 1$^+$ intermediate nucleus. Therefore, although we conceive of double-beta decay as a sequence of two $\beta^+$, $\beta^-$ or electron capture decays \cite{DeOliveira2015}, in the present study we calculate $\beta^-$ decay from the initial 0$^+$ nucleus and a $\beta^+$ decay from the final 0$^+$ nucleus, resulting in two sets of intermediate 1$^+$ states. Explicitly, for the GT transitions the matrix element is given by:

\begin{equation}
\mathcal{M}{^{2\nu}_{GT}} =  m_e \sum_{m'm} \frac{ \bra{m}\sum_a \sigma_a \tau^-_a\ket{i}\braket{m|m'}\bra{f}\sum_a \sigma_a \tau^-_a\ket{m'} }{E_{m}+\frac{Q}{2}+m_e}.
\end{equation}

For the overlap between states belonging to different sets we take the usual prescription \cite{Simkovic1997}:

\begin{equation}
\label{eq5}
\braket{m|m'} \approx \sum_{pn} \left[ X^m_{pn} X^{m'}_{pn} - Y^m_{pn} Y^{m'}_{pn} \right],
\end{equation}

where the quantities X and Y are the REDF-QRPA amplitudes that will be defined in the next section. This is an approximation in cases where the initial and final states are not identical, but a reasonable one. For $^{48}$Ca, for example, we calculate the overlap between the lowest states belonging to each of the different sets of states in the intermediate nucleus to be 0.9855. Higher states in one set are more likely to have significant overlap with several states in the other, but the values of the overlaps still go from around 0.14 toward higher ones around 0.89. An alternative prescription from \v Simkovic~\cite{Simkovic1997}, involves additional factors proportional to occupation numbers:

\begin{equation}
\label{eq6}
\braket{m|m'} \approx \sum_{pn} \left[ X^m_{pn} X^{m'}_{pn} - Y^m_{pn} Y^{m'}_{pn} \right]\tilde{u}_p\tilde{u}_n,
\end{equation}

where the quantities $\tilde{u}_{p/n}$ are defined as follows:

\begin{equation}
\label{eq7}
\tilde{u}_{p/n} = u_{p/n}^{m}u_{p/n}^{m'} + v_{p/n}^{m}v_{p/n}^{m'},
\end{equation}

and $u_{p/n}$, $v_{p/n}$ are occupation numbers derived, in the present work, from the relativistic Hartree-BCS model (more details are given in Sec. \ref{relativistic}).
Several effects are usually neglected in the study of 2$\nu\beta\beta$ decays,
such as higher-order currents and realistic short-range correlations \cite{Vogel2012},
 that are important in the neutrinoless case. This is due to the insensitivity of the two-neutrino NMEs to the details of the nucleon wavefunctions for low nucleon separation r$_{12}$~\cite{Vogel2012}.

\section{Relativistic framework for $2\nu\beta\beta$ decay matrix elements \label{relativistic}}
\subsection{Relativistic Hartree-BCS for the ground state description}\label{secrhb}
Calculations of nuclear matrix elements for double-beta decay often proceed starting from a realistic nucleon-nucleon potential \cite{Aunola1996,Vergados2016}. As mentioned in
Sec. \ref{Intro}, other theory frameworks based on phenomenological effective interactions have also been employed, in particular, shell model and nuclear energy density functionals. In this work we introduce a framework for 2$\nu\beta\beta$ decays based on a relativistic nuclear energy density functional \cite{nik14}. 
In the relativistic nuclear energy density functional (REDF) framework, the nuclear ground-state
density and energy are determined by the self-consistent solution of relativistic single-nucleon
Kohn-Sham equations~\cite{koh1,koh2}. In the present study these equations are implemented 
through an interaction Lagrangian density formulated in terms of the relevant degrees of freedom.  Since the REDF has already been extensively used in many previous studies, here we give only a brief overview of the relevant foundations of this framework.
Two different families of relativistic density-dependent interactions are used in this work, (i) finite range meson-exchange and (ii) point coupling interactions. In the former case, 
pointlike nucleons interact through the exchange of light mesons, namely $\omega$, $\rho$ and $\sigma$ mesons, in addition to an electromagnetic interaction mediated by photons. The model is explained in detail in \cite{Lalazissis2005,Hofman2001,Typel1999,Fuchs1995,nik14}.
Meson-nucleon couplings are established as functions of the vector density, motivated by the relativistic Brueckner-Hartree-Fock calculations, but introduced through a phenomenological 
ansatz with parameters adjusted to the experimental data in finite nuclei \cite{nik14}. In this work, the DD-ME2 parameterization of the density-dependent meson-exchange interaction is used \cite{Lalazissis2005}, being one of the most successful parameterizations currently used in the description of a variety of nuclear properties and astrophysically relevant processes \cite{Paar2007,Paar2015,RocaMaza2018}.

In the case of point coupling interactions, the effective Lagrangian contains four fermion contact interaction terms including the isoscalar-scalar, isoscalar-vector, isovector-vector and isospace-space channels, coupling of protons to the electromagnetic field, and the derivative term accounting for the leading effects of finite-range interactions necessary for a quantitative description of nuclear density distribution and radii (for more details see Refs.~\cite{Nik08,nik14}). In our study of 2$\nu\beta\beta$ decays, two parameterizations of the density dependent point coupling interactions are used, DD-PC1 \cite{Nik08}, and the more recently established DD-PCX \cite{Yuksel2019}. While the DD-PC1 interaction is adjusted to
nuclear binding energies, the DD-PCX interaction is specifically adjusted both to the nuclear ground state and excitation properties, in order to constrain
not only the nuclear properties but also the symmetry energy close to the saturation density, and the incompressibility of nuclear matter by using genuine observables on finite nuclei in the $\chi^2$ minimization \cite{Yuksel2019}.

For the description of ground state properties of open-shell nuclei a unified and self-consistent treatment of mean-field and pairing correlations is needed. In this work we employ the relativistic Hartree-BCS (RH-BCS) theory \cite{Ravlic2021,Yuksel2020} 
which represents a relativistic extension of the non-relativistic Hartree-Fock-BCS framework \cite{Bonche2005,Reinhard2021}. Spherical symmetry is assumed.
The pairing correlations in the ground state are described using a separable 
pairing force, which also includes two parameters for the pairing strength ($G_{p}$ and $G_{n}$)
\cite{Tian2009}. While in the RH-BCS model with DD-ME2 and DD-PC1 interactions the
pairing parameterization from Ref. \cite{Tian2009} is used, the DD-PCX interaction is supplemented
with its own parameterization for the separable pairing force as given in Ref. \cite{Yuksel2019}.
Within the  RH-BCS model, the isoscalar pairing is not included, i.e., no proton-neutron mixing is considered.

\subsection{Proton-neutron relativistic QRPA\label{secqrpa}}

Charge-exchange transitions between the states in nuclides involved in $2\nu\beta\beta$ decay are described in the framework of the proton-neutron relativistic quasiparticle random phase approximation (REDF-QRPA). An introduction to the charge-exchange QRPA (or proton-neutron QRPA, pn-QRPA) calculations can be found in Refs.~\cite{Bai2014,Toivanen1995}, while in the relativistic framework it has been introduced in Refs.~\cite{Paar2003,Paar2004}. 
The residual REDF-QRPA interaction is derived from the relativistic formulation of the effective Lagrangian 
density, and throughout the calculation of the QRPA matrix elements, the Dirac wave functions from the RH-BCS model,
including both large and small components are systematically included. In
addition, the transition operators are also extended for their implementation
in the relativistic framework. More details on the relativistic QRPA are given in Refs.~\cite{Paar2003,Paar2004,Paar2007,Daoutidis2011,Niksic2013}.
Here we give only a brief overview of the REDF-QRPA adopted for the purpose of the study of $2\nu\beta\beta$ decay. The states in the intermediate $(Z+1,N-1)$ nucleus are REDF-QRPA phonons:
\begin{equation}
\ket{m} = \sum_{pn}\left[ X_{pn}a^{\dagger}_p a^{\dagger}_n - Y_{pn}a_n a_p \right]\ket{\text{QRPA}},
\end{equation}
where the creation operators $a^\dag_{p(n)}$ create a proton(neutron) state in an orbital labeled $p(n)$, and quantities $X$ and $Y$ are, as noted earlier, the REDF-QRPA amplitudes. $\ket{\text{QRPA}}$ denotes the QRPA vacuum, which we take to be the ground state (see Sec.\ref{secrhb}). By linearizing the time-dependent RH-BCS equations in charge-exchange external field, the charge-exhange QRPA  equations are obtained \cite{Suhonen2007}:
\begin{equation}
\begin{pmatrix}
A & B \\ -B^* & -A^*
\end{pmatrix} 
\begin{pmatrix}
X(J) \\ Y(J)
\end{pmatrix} = \omega_k
\begin{pmatrix}
X(J) \\ Y(J)
\end{pmatrix},
\end{equation}
where the REDF-QRPA matrices $A$ and $B$ are defined as
\begin{align}
\begin{split}
 &A_{pnp'n'}(J) = (E_p+E_n)\delta_{pp'}\delta_{nn'}  \\ 
 &+ (u_pu_{p'}u_nu_{n'}+v_pv_{p'}v_nv_{n'})\bra{pnJ}V\ket{p'n'J} \\ 
 &+  (u_pv_{p'}u_nv_{n'}+v_pu_{p'}v_nu_{n'})\bra{pn^{-1}J}V_{res}\ket{p'n'^{-1}J},
 \end{split}
\end{align}
and
\begin{align}
\begin{split}
 &B_{pnp'n'}(J) =  (u_pu_{p'}v_nv_{n'}+v_pv_{p'}u_nu_{n'})\bra{pnJ}V\ket{p'n'J}  \\
 &+  (u_pv_{p'}v_nu_{n'}+v_pu_{p'}u_nv_{n'})\bra{pn^{-1}J}V_{res}\ket{p'n'^{-1}J},
 \end{split}
\end{align}
where $V_{res}$ is the residual interaction derived from the relativistic nuclear energy density functional, while $V$ includes the pairing interaction in the QRPA \cite{Paar2004,Vale2020}, $u_{p(n)}, v_{p(n)}$ being proton(neutron) RH-BCS amplitudes.
The residual interaction  $V_{res}$  is derived from the same effective meson-exchange or point coupling interaction as used in the ground state calculations (see Sec. \ref{secrhb} ). In addition, the pseudovector interaction 
channel is also included, and its strength parameter has been previously constrained to
the experimental data on Gamow-Teller resonance in $^{208}$Pb \cite{Paar2004,Paar2007,Vale2020}.

The REDF-QRPA equations include both the isovector $(T=1)$ and isoscalar $(T=0)$ pairing channels, described by the separable interaction \cite{Tian2009,Vale2020,PhysRevC.104.064302},
\begin{equation}
V^{pn}(r_1,r_2) = -f G\delta(\boldsymbol{R}-\boldsymbol{R}^\prime)P(r)P(r^\prime)(1-P^r P^{\sigma}P^{\tau}),
\end{equation}
where the projectors $P^{r,\sigma,\tau}$ are defined as usual, $G$ is the overall interaction strength, $R$ and $r$ are center of mass and relative coordinates respectively, and $P(r)$ is defined as:
\begin{equation}
P(r) = \frac{1}{(4\pi a^2)^{3/2}e^{-\frac{r^2}{2a^2}}}.
\end{equation}
The overall strength in the pairing channel is multiplied by a dimensionless
factor $f$ defined as
\begin{equation}
f = \left\{ \begin{array}{c}
1, \quad T = 1, S = 0, \\
V_{0pp}, \quad T=0, S = 1, \\
0, \quad \text{ otherwise},
\end{array} \right. 
\end{equation}
where $V_{0pp}$ represents the isoscalar pairing strength. 
For $T=1$  channel, the same separable pairing interaction is used as in the RH-BCS model (see Sec. \ref{secrhb}). The strength parameters $G_{p(n)}$ for protons(neutrons) and the width $a$ are defined in Ref. \cite{PhysRevC.79.064301}, but other parameterizations also exist, e.g., the one constrained with the DD-PCX interaction \cite{Yuksel2019}. For $T=0$  channel, the pairing strength parameter $V_{0pp}$ has to be constrained at the level of  REDF-QRPA, e.g., by using experimental data on Gamow-Teller transitions or $\beta$-decays.
In our treatment of 2$\nu\beta\beta$ decay, this allows us to explore the dependence
of the NME on the pairing strength parameter $V_{0pp}$, showing the variation in the possible values for the matrix elements. However, to provide predictions on 2$\nu\beta\beta$ decay
matrix elements, in this work $V_{0pp}$ is also constrained by the experimental data on single $\beta$-decay
half-lives (Sec. \ref{optimizeV0}).

%
The transition matrix elements necessary for 2$\nu\beta\beta$  NME in Eq. (\ref{eq2}) are given by following
expressions \cite{Vogel1986},
\begin{align}
\begin{split}
&\bra{i}\sum_a \sigma_a \tau^-_a \ket{m}  \\
&= \sum_{pn} \langle p || \boldsymbol{\sigma} ||n \rangle \left[ u_p v_n X^m_{pn} + v_p u_n Y^m_{pn} \right] \\
\end{split}
\end{align}
\begin{align}
\begin{split}
&\bra{m}\sum_a \sigma_a \tau^+_a \ket{f}  = \\
 &\sum_{pn} \langle p || \boldsymbol{\sigma}  || n \rangle \left[ v_p u_n X^m_{pn} + u_p v_n Y^m_{pn} \right].
\end{split}
\end{align}
Here, the terms next to the $X$ and $Y$ amplitudes represent particle-type and hole-type one-quasi-particle transitions, respectively \cite{Suhonen2007}. 

\subsection{Determination of the isoscalar pairing strength\label{optimizeV0}}

In order to constrain the isoscalar ($T = 0$) pairing strength $V_{0pp}$ we use similar approach as suggested in Refs. \cite{Ravlic2021,Marketin2016,NIU2013172}, namely $V_{0pp}$ is determined from the global fit of $\beta$-decay half-lives to experimental data. We have used even-even nuclei in the range $8 \leq Z \leq 82$ for which experimental data on $\beta$-decay half-lives is available,
where $V_{0pp}$ has the effect in the relative change of half-lives $T_{1/2}$ by more than 20$\%$, and
whose half-lives are $<\text{10}^3$ s. In this way we have additionally optimized the fitting procedure, because too long half-lives may be challenging for quantitative description within the QRPA method \cite{Marketin2016}. Since the isoscalar pairing strength $V_{0pp}$ shows isotopic dependence \cite{Marketin2016}, in the fitting procedure we use a functional form,
\begin{equation}
V_{0pp}=V_1+V_2\left(\frac{N-Z}{A}\right)
\label{v0eq}
\end{equation}
that appears to provide a comparable quality of the fit to $\beta$-decay half-lives to 
other previously used functional forms \cite{Marketin2016}.
The $\beta$-decay half-lives are calculated using the REDF-QRPA as described in Ref. \cite{Ravlic2021}, including both allowed and first-forbidden transitions. We have used the same quenched value of axial-coupling $g_A=1.0$ as in this work. The fitting procedure determines $V_{0pp}$ which reproduces the experimental $\beta$-decay half-lives. Here we note that only those $V_{0pp}$ which yield real solutions of the QRPA equation should be included in the fit to the proposed ansatz.
Using models with DD-ME2, DD-PC1 and DD-PCX interactions, the obtained average values of the fitted parameters $V_1$ and $V_2$ as well as $1\sigma-$uncertainties are given in Tab. \ref{fitpar}.
This results in rather narrow range of values for the isoscalar pairing strength $V_{0pp}$ for the set of nuclei 
considered in 2$\nu\beta\beta$ decay study in this work. The respective values for $V_{0pp}$ with 
the uncertainties are given in Tab. \ref{tableV0pp}. In the following Section we refer to \emph{optimal} $V_{0pp}$ when using the values from Tab. \ref{tableV0pp}.
{We note that most of the $V_{0pp}$ values are close to 1.0, meaning the strengths of the isoscalar and isovector pairing are very similar, which in turns points to a "soft" breaking of spin-isospin SU(4) symmetry 
in present calculations.}

\begin{table}

\caption{The fit parameters for the $T=0$ pairing strength functional form in Eq. (\ref{v0eq}) obtained from the optimization on $\beta$-decay half-lives for the models with DD-PCX, DD-PC1 and DD-ME2 interactions} \label{fitpar}
\begin{ruledtabular}
\begin{tabular}{ cccc }
& DD-ME2 & DD-PC1 & DD-PCX\\
\hline 
$V_1$ & 0.574 $\pm$ 0.338 & 0.522 $\pm$ 0.407	&  0.592 $\pm$ 0.366 \\
$V_2$ & 2.301 $\pm$ 1.422 & 3.092 $\pm$ 1.768	&  2.321 $\pm$ 1.559 \\
\end{tabular}
\end{ruledtabular}
\end{table}

\begin{table}
\caption{Optimal values of the $T=0$ pairing strength parameter $V_{0pp}$ constrained
from the $\beta$-decay half lives.\label{tableV0pp}}
\begin{ruledtabular}
\begin{tabular}{ cccc }
& & $V_{0pp}$ \\
& DD-ME2 & DD-PC1 & DD-PCX\\
\hline 
$^{48}$Ca & 0.98 $\pm$ 0.12 & 1.04 $\pm$ 0.12	&  0.96 $\pm$ 0.11 \\
$^{76}$Ge & 0.96 $\pm$ 0.13 & 1.01 $\pm$ 0.14	&  0.94 $\pm$ 0.12 \\
$^{82}$Se & 0.99 $\pm$ 0.11 & 1.05 $\pm$ 0.12	&  0.97 $\pm$ 0.10 \\
$^{96}$Zr & 0.98 $\pm$ 0.12 & 1.04 $\pm$ 0.12	&  0.96 $\pm$ 0.11 \\
$^{100}$Mo &0.96 $\pm$ 0.13 & 1.02 $\pm$ 0.13	&  0.94 $\pm$ 0.12 \\
$^{116}$Cd &0.99 $\pm$ 0.11 & 1.06 $\pm$ 0.11	&  0.97 $\pm$ 0.10 \\
$^{124}$Xe &0.89 $\pm$ 0.17 & 0.92 $\pm$ 0.18	&  0.87 $\pm$ 0.16 \\
$^{128}$Te &1.07 $\pm$ 0.07 & 1.16 $\pm$ 0.07	&  1.05 $\pm$ 0.07 \\
$^{130}$Te &1.03 $\pm$ 0.09 & 1.10 $\pm$ 0.09	&  1.01 $\pm$ 0.08 \\
$^{136}$Xe &1.06 $\pm$ 0.08 & 1.14 $\pm$ 0.08	&  1.03 $\pm$ 0.07 \\
$^{150}$Nd &1.06 $\pm$ 0.08 & 1.14 $\pm$ 0.08	&  1.03 $\pm$ 0.07 \\
\end{tabular}
\end{ruledtabular}
\end{table}

\section{Results and discussion}

By employing the framework of a relativistic theory for nuclear properties and transitions as outlined
in the previous sections, we have performed calculations of the $2\nu\beta\beta$ decay matrix elements. In the first step, the RH-BCS model \cite{nik14} has been used to describe the ground state properties of the initial and final nuclei involved in the decay. The RH-BCS model is formulated in the harmonic oscillator basis and we restrict calculations to 20 oscillator shells 
both for protons and neutrons. The single-particle wave functions and the corresponding occupation probabilities in the RH-BCS quasiparticle basis are used in the REDF-QRPA to describe 
Gamow-Teller (GT) transitions involved in the  $2\nu\beta\beta$ decay.
The REDF-QRPA calculations are performed in two steps, for beta minus (plus) matrix elements for decays from the initial (final) nucleus to the intermediate nucleus, which are then used to calculate the 2$\nu\beta\beta$ decay NME. All the other quantities that appear in the calculations, including the energies of the intermediate states and the $Q$-values, are also taken self-consistently from the RH-BCS and REDF-QRPA calculations. In order to assess the information on the model dependence of the 
2$\nu\beta\beta$ decay NME, three relativistic energy density functionals are used
in the study, including the density-dependent meson-exchange interaction with the DD-ME2 parameterization \cite{Lalazissis2005}, and density dependent  point coupling 
interactions  DD-PC1 \cite{Nik08}, and the more recently established 
DD-PCX \cite{Yuksel2019}. 
One of the open questions in the description of double-beta decays is the problem of quenching of the axial-vector coupling constant $g_A$, that has attracted attention in many recent studies, e.g. see Refs. \cite{Mustonen2013,Engel2017}. As has already been mentioned above, $g_A=1.0$ is systematically 
used in the present study.

\begin{figure}[h!]
\includegraphics[scale=0.4]{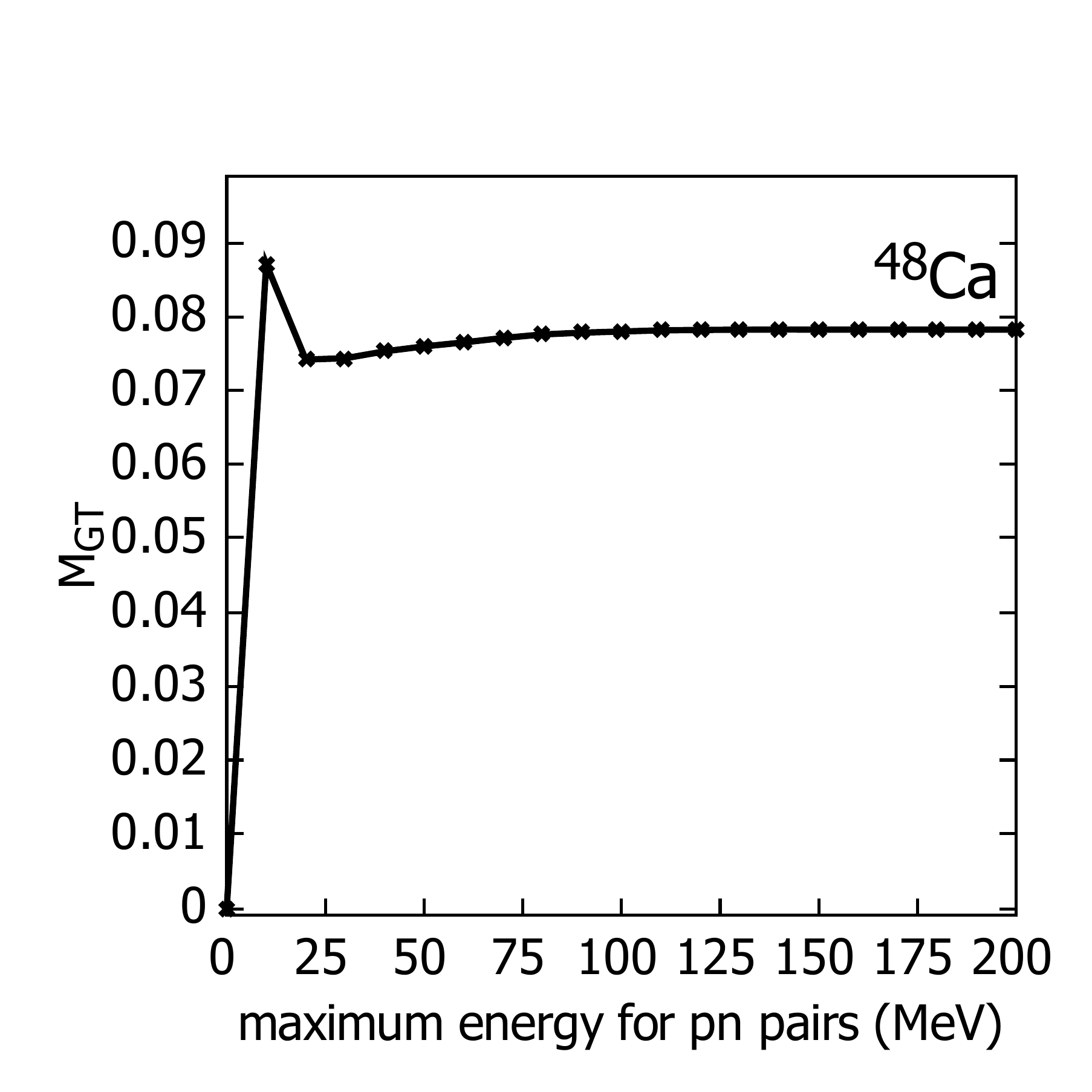}
\caption{The NMEs for $2\nu\beta\beta$ decay based on Gamow-Teller transitions, shown as a function of the maximal two-quasiparticle energy. The DD-ME2 interaction is used in calculations. Matrix elements are dimensionless.}
\label{fig1}
\end{figure}

Throughout this section we provide the absolute values of the $2\nu\beta\beta$ decay matrix elements, since only the squares of their absolute values have physical significance.
The REDF-QRPA calculations are first performed to set the cut-off energy for the 
quasiparticle pairs that compose the configuration space. To all relevant
configurations, a convergence test is performed in order to restrict the maximal
two-quasiparticle energy with the condition that the value of the NME converges with
increasing energy. We illustrate the convergence of the NME values in Fig. \ref{fig1}
for the case of $^{48}$Ca, where the matrix elements are shown as a function of
the maximal proton-neutron two-quasiparticle energy $E_{2qp}^{max}$. The DD-ME2 effective
interaction is used for this demonstration.
The Gamow-Teller (GT) transitions
are considered. One can observe that the NMEs 
converge with high accuracy at two-quasiparticle energy $E_{2qp}^{max}\approx$ 100 MeV.
%
%

Insight into the contributions of various states to the final NMEs can be obtained from the running sum, that includes the sum of all contributions to the matrix element up to a specific maximal excitation energy in the intermediate nucleus~\cite{Guerra2012} which we denote as E$_{exc}$ in the following discussion. 

\begin{figure}[h!]
\includegraphics[scale=0.4]{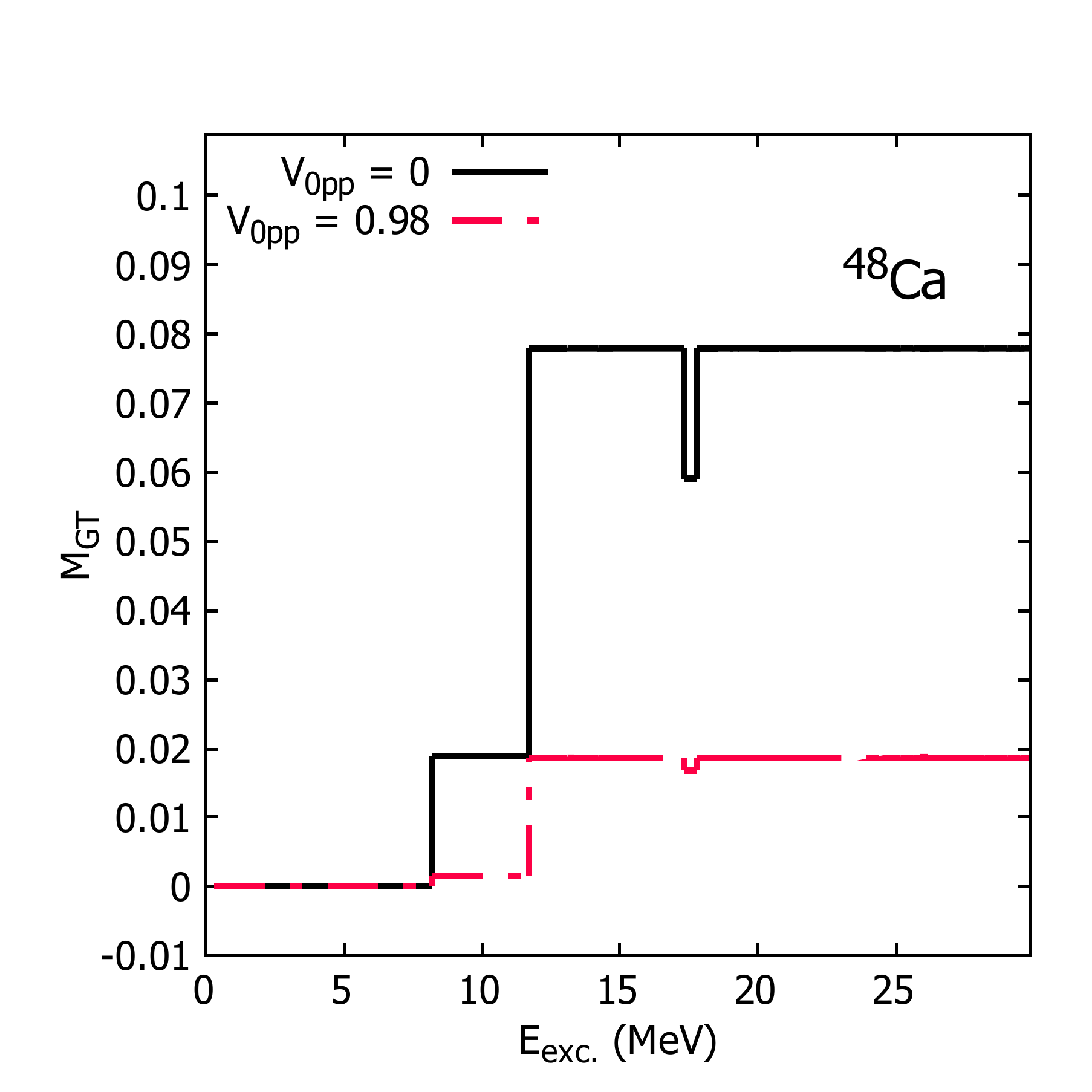}
\caption{The running sum of the GT NMEs for the $2\nu\beta\beta$ decay of $^{48}$Ca for the DD-ME2 interaction, shown as a function of maximal excitation energy E$_{exc}$ in the intermediate nucleus. The cases with and without $T=0$ pairing are shown separately. \label{fig2}}
\end{figure}

\begin{figure}[h!]
\includegraphics[scale=0.4]{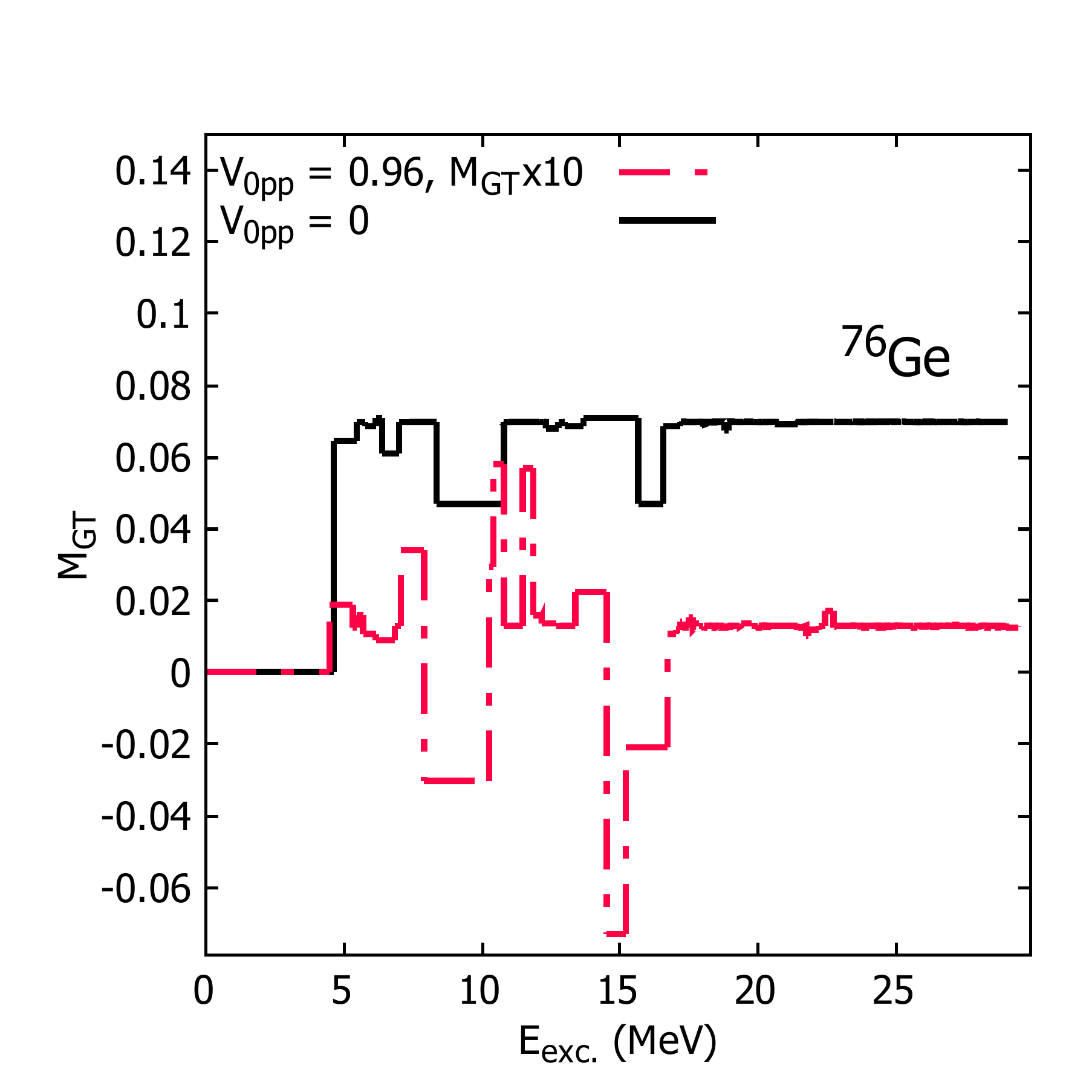}
\caption{The same as in Fig.\ref{fig2}, but for  $^{76}$Ge.
\label{fig3}}
\end{figure}

\begin{figure}[h!]
\includegraphics[scale=0.4]{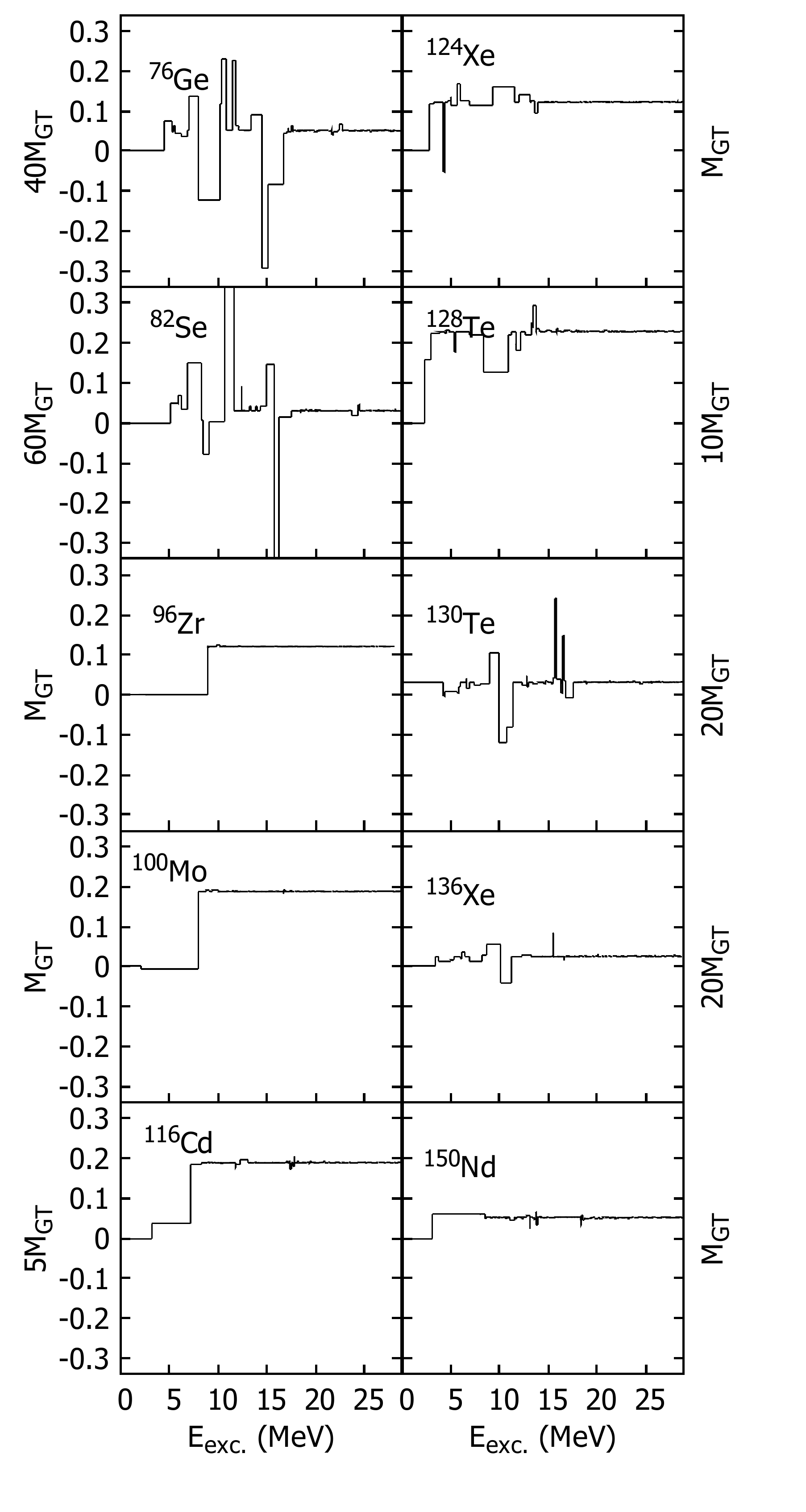}
\caption{Running sums for the GT NMEs for DD-ME2 interaction and optimal values of $V_{0pp}$ from Tab. \ref{tableV0pp}, for the $2\nu\beta\beta$ decays of $^{82}$Se--$^{150}$Nd, shown as a function of maximal excitation energy E$_{exc}$ in the intermediate nucleus.
\label{fignn1}}
\end{figure}

Figures \ref{fig2} and \ref{fig3}
show the respective running sums for the Gamow-Teller double-beta decay transitions
for $^{48}$Ca and $^{76}$Ge, displayed as functions of E$_{exc}$. The sums are taken at two values of the $T=0$ pairing strength parameter, $V_{0pp}= 0$, representing a situation where isoscalar pairing vanishes, and the optimal values $V_{0pp}=0.98$ and 0.96 for $^{48}$Ca and $^{76}$Ge, respectively.
One can observe that for $^{48}$Ca the hypothesis of single state dominance \cite{Simkovic2001} holds in the REDF-QRPA calculations, i.e., most of the contribution to the NMEs mainly comes from a single low-lying state in the intermediate nucleus. For $^{76}$Ge, many states contribute to $M_{GT}$ up to 20 MeV, and in some cases destructive interference in their contributions to the NMEs is obtained.
The same dependence is shown in Fig. \ref{fignn1} for other nuclei considered in this work, from $^{82}$Se to $^{150}$Nd, using optimal values of $V_{0pp}$. We see that generally the single-state dominance (SSD) hypothesis is approximatively fulfilled, although in some cases more complicated structure
is obtained due to contributions from several states. Particularly noticeable are significant cancellations when the nuclear matrix element is significantly lowered from the value at zero isoscalar pairing due to a high value of V$_{0pp}$.
It is interesting to note that presented results {for some nuclei} are in reasonable agreement with recent experimental results, which show that the SSD hypothesis holds for $^{100}$Mo~\citep{Azzolini2019}, while for  $^{82}$Se we obtain a more complex structure than the SSD measured in Ref. \cite{Coraggio2019}. {Our results for the running sum in $^{136}$Xe are at variance with the experimental results in Ref. \cite{Gando2019}.
In the same study, it has been shown that the running sum is strongly model
dependent, resulting in considerable differences between the reported QRPA and shell
model calculations \cite{Gando2019}.}
{Our results also differ from other calculations of running sums, including
also contributions with a negative sign. They appear and become more prominent with increasing V$_{0pp}$, and cancel out 
to a large extent an initially large NME, and the fact that we show the running sums at higher optimal V$_{0pp}$ value explains in part their appearance.}

\begin{figure}[h!]
\includegraphics[scale=0.4]{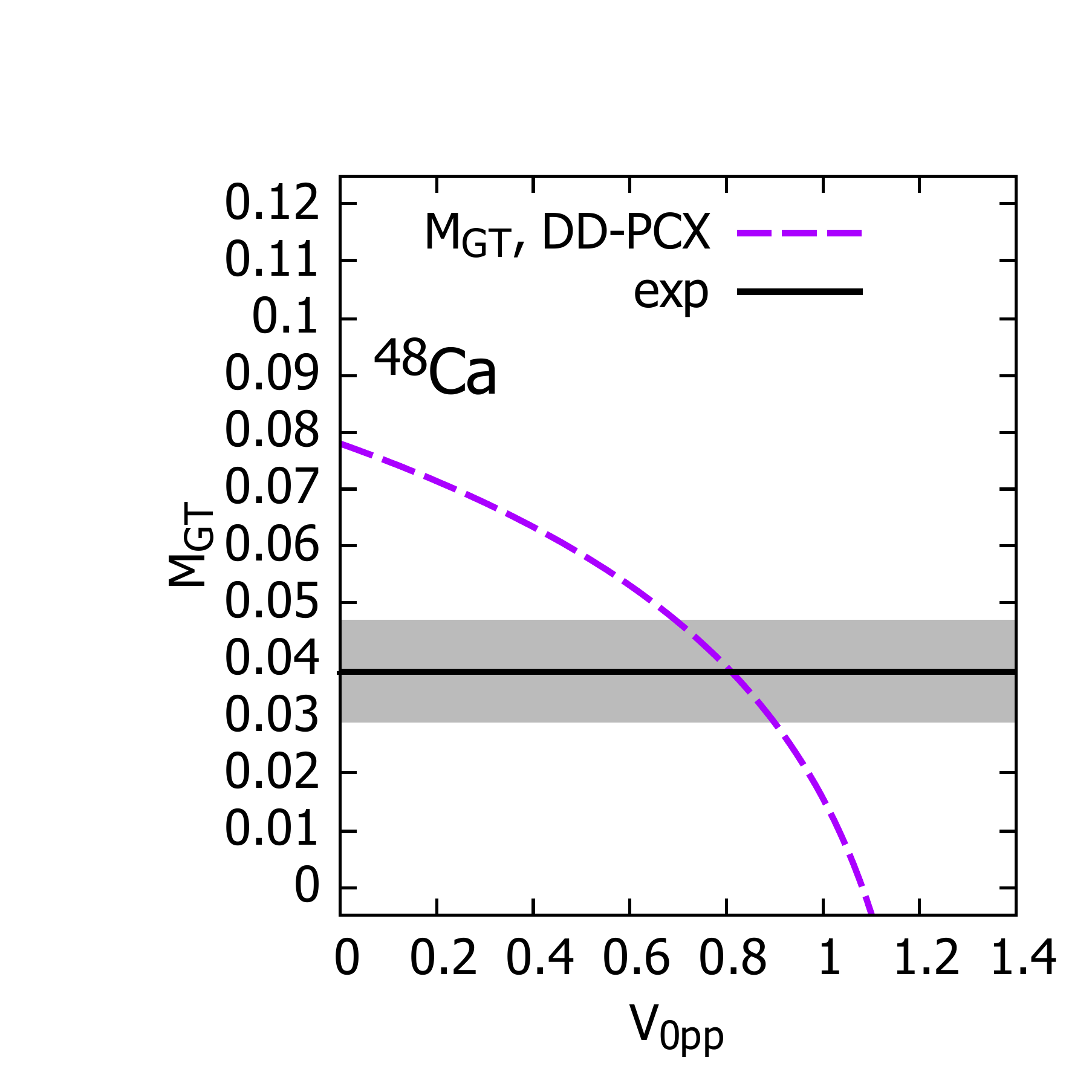}
\caption{The dependence of the GT NMEs for 2$\nu\beta\beta$ decay
on the isoscalar pairing strength $V_{0pp}$ for $^{48}$Ca, using DD-ME2 interaction, in comparison to the result obtained from the experimental data on $2\nu\beta\beta$ decay \cite{Barabash20152}. 
\label{fig4}}
\end{figure}

Before systematic implementation of the optimal values of $V_{0pp}$ from Tab. \ref{tableV0pp} in 
model calculations, it is interesting to explore the general dependence and sensitivity of the $2\nu\beta\beta$ decay 
NMEs on $V_{0pp}$. In Fig. \ref{fig4}
the dependence of the absolute values of nuclear matrix elements is shown for $^{48}$Ca 
as a function of $V_{0pp}$. The black line in figure represents the value of the nuclear matrix element deduced from the experimental data on $2\nu\beta\beta$ half-lives \cite{Barabash20152,Barabash2015}. The grey band around this line denotes an uncertainty at the level of 3$\sigma$. 
As one can see in Fig. \ref{fig4}, the NME systematically decreases
with $V_{0pp}$, and an overlap between the calculated curve and the experimental limit 
for the NMEs is obtained.

\begin{figure}[h!]
\includegraphics[scale=0.4]{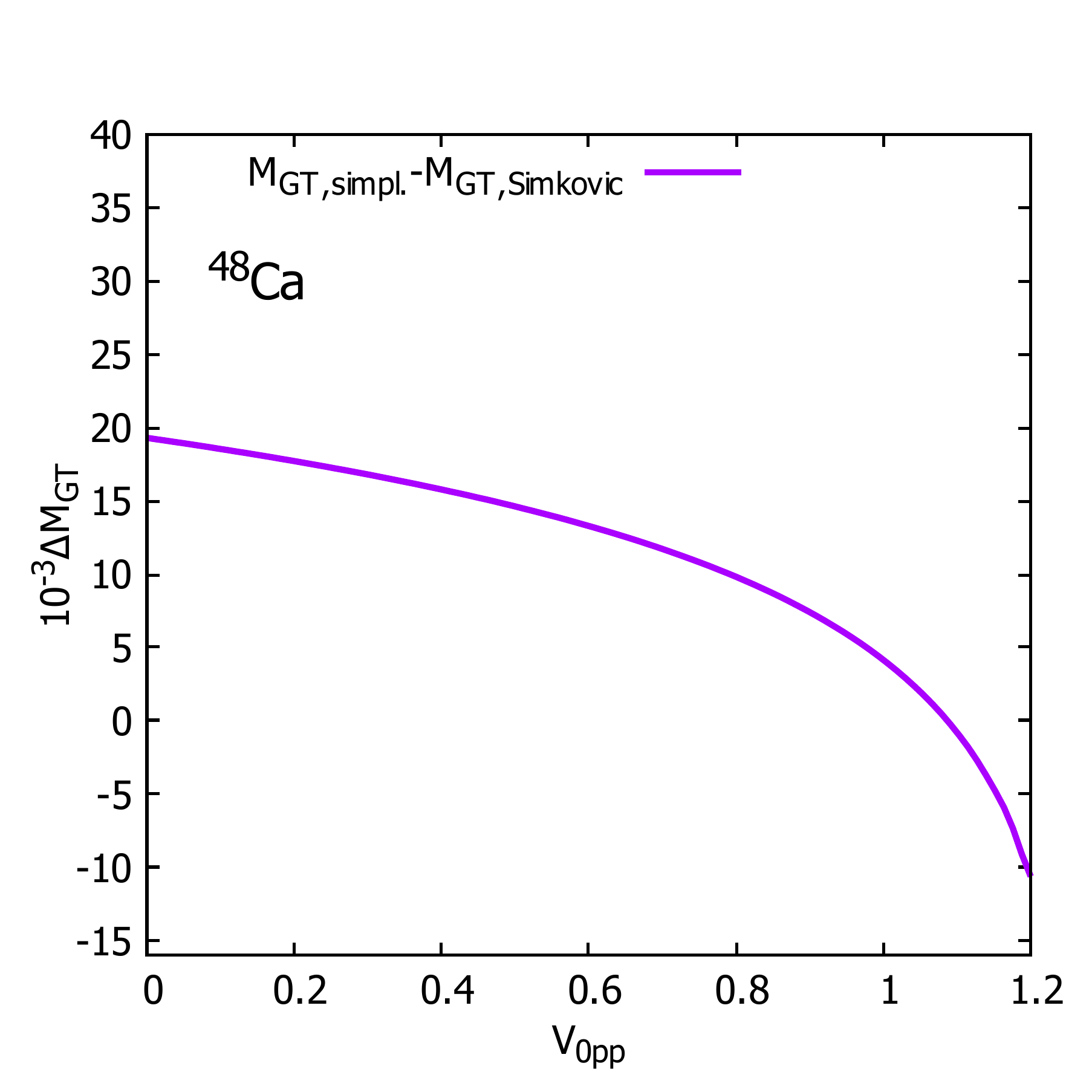}
\caption{The difference between the NMEs calculated using two prescriptions for the GT $2\nu\beta\beta$ decay matrix element, shown as a function of the isoscalar pairing strength $V_{0pp}$ (see text for details). 'M$_{GT,simpl.}$' refers to the usual prescription for the overlap factor in Eq. (\ref{eq5}), while 'M$_{GT,Simkovic}$' refers to the prescription in Eq.(\ref{eq6}).\label{fig5}}
\end{figure}

Here, the overlap prescription given in Eq. (\ref{eq5}) is used. 
We have verified that the result using the overlap prescription in Eq. (\ref{eq5}) appears almost identical to the one given in Eq. (\ref{eq6}). 
Figure \ref{fig5} shows the difference between the NMEs for $^{48}$Ca obtained using these two methods for calculating the overlap, given as a function of $V_{0pp}$.
One can observe that not only the difference is small, but it also reduces as one approaches the 
$V_{0pp}$ value deduced from the experiment as the optimal one, as shown in Fig. \ref{fig4}.
Therefore, in the following investigation we consider only the results based on the overlap
prescription given in Eq. (\ref{eq5}).

\begin{figure}[h!]
\includegraphics[scale=0.4]{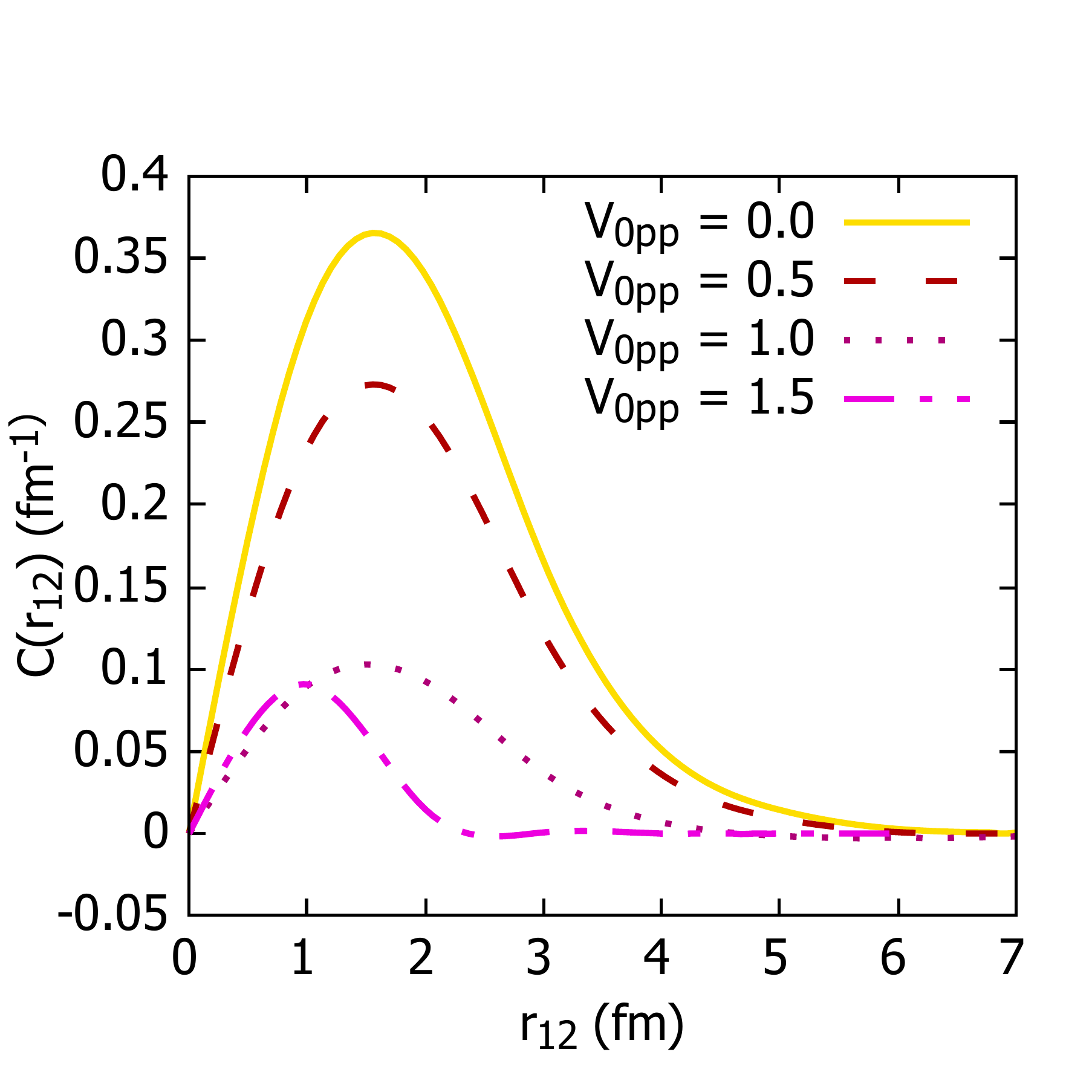}
\caption{The function $C(r_{12})$, showing the contributions to the NME at nucleon separations $r_{12}$, for the Gamow-Teller $2\nu\beta\beta$ decay of $^{48}$Ca, for $V_{0pp}$ from 0.0 to 1.5.\label{figC}}
\end{figure}

A quantity related to the NMEs that has received recent theoretical and experimental interest~\cite{Shimizu2018,Simkovic2008} is the function $C(r)$, representing the contribution to the NME at a given internucleon distance $r_{12}\equiv r$. It is related to the NME, denoted here as $\mathcal{M}$, as

\begin{equation}
\int_0^\infty d r C(r) = \mathcal{M}.
\end{equation}
Figure \ref{figC}  shows the $C(r)$ function for 
2$\nu\beta\beta$ decay of $^{48}$Ca, obtained using the DD-ME2 interaction for the range of values of $T=0$ pairing strength parameters $V_{0pp}$. The quantity is evaluated in the closure approximation and is directly connected to the dimensionless NME. It needs to be noted that, due to the computational resources necessary for the evaluation of the $C(r)$ function, the results we report have been calculated at a lower energy cutoff than the results for the NMEs in the present work, specifically at 50 MeV. Comparing the $C(r)$ function to those calculated in Ref.~\cite{Simkovic2008} we conclude that qualitatively the same shape is obtained, with a sharp peak around 1 fm, converging to zero with an increase of $r$.
This result is consistent with the findings that this shape is universal for 2$\nu\beta\beta$ decay \cite{Menendez2009}, although our model does not contain $S=0$ pairing in the $T=0$ channel, as well as $T=0$ channel in the ground state. The behaviour obtained with increasing $T=0$ pairing strength $V_{0pp}$ in the residual QRPA interaction is also consistent with previous studies \cite{Simkovic2008}, i.e., the $C(r)$ function reaches its highest values when pairing is not taken into account and with the increase of the pairing strength the central peak becomes reduced. 
We would note that the contribution of high inter-nucleon distances in many nuclei is negative and serves to lower the NME; these negative contributions, once again, become more noticeable with increasing V$_{0pp}$.

\begin{figure}[h!]
\includegraphics[scale=0.4]{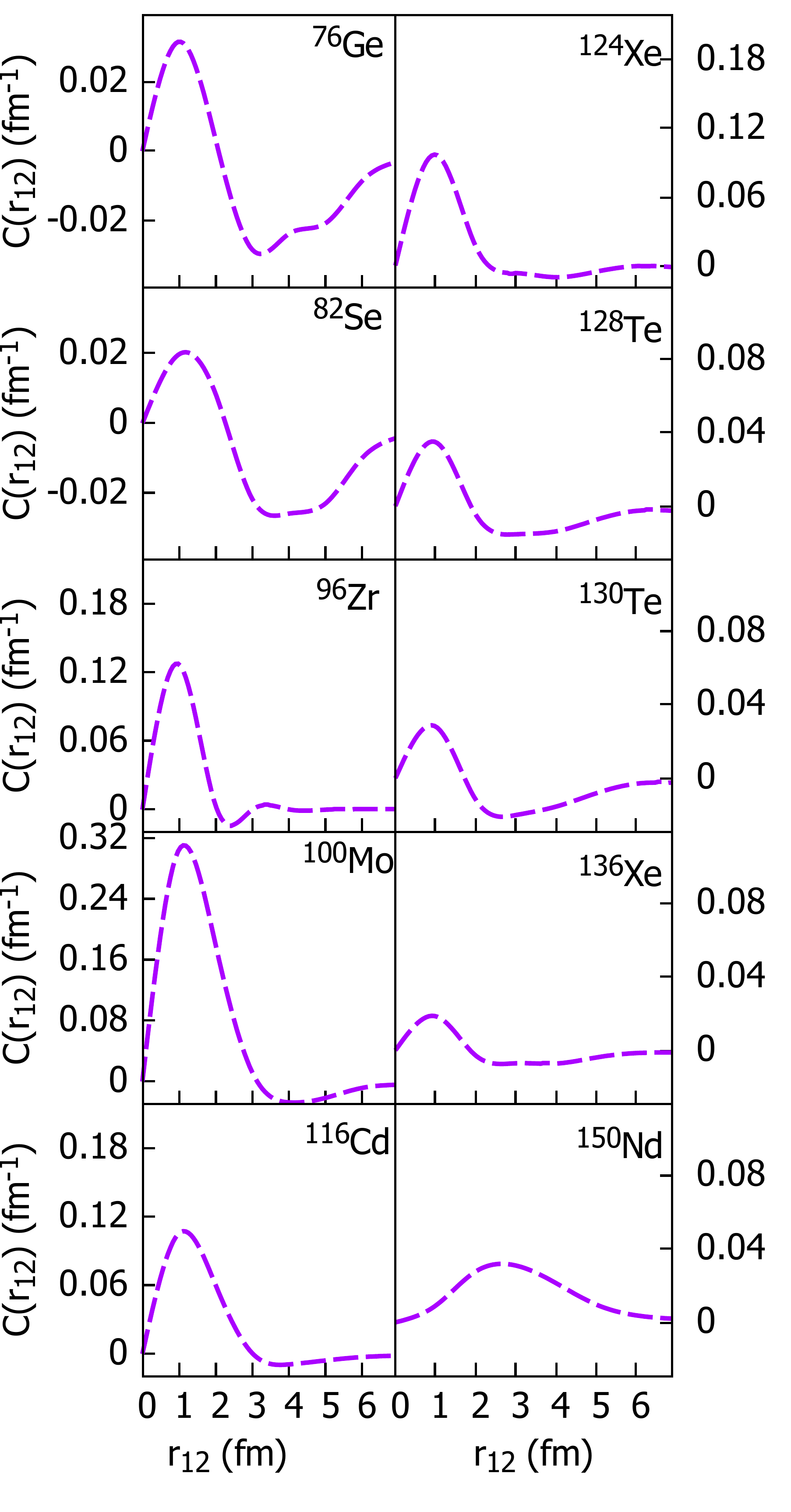}
\caption{The function $C(r_{12})$, showing the contributions to the NME at nucleon separations $r_{12}$, for the Gamow-Teller $2\nu\beta\beta$ decay of $^{76}$Ge-$^{150}$Nd, evaluated at optimal values of $V_{0pp}$.\label{figC2}}
\end{figure}

The same function, for other nuclides considered in this work up to $^{150}$Nd, evaluated using DD-ME2 interaction and optimal values of $V_{0pp}$, is shown in Fig. \ref{figC2}. Similar general behavior of $C(r_{12})$ function is obtained for all nuclei, 
{that peaks strongly at low $r_{12}$ values. Where the NMEs are close to zero, 
we can note significant cancellations as one integrates the $C(r)$ function over the inter-nucleon distance r.
$^{150}$Nd is an exception to the general rule that the first and most prominent peak of the function is near 1 fm, with a dominant peak at significantly higher radii. This suggests that the results for $^{150}$Nd should be taken with less certainty than those for other nuclides.
However, we need to point out that in certain cases our C(r) functions also show contributions at higher inter-nucleon distances than is found in previous
QRPA calculations. We note that the contribution of high inter-nucleon 
distances in many nuclei is negative and serves to lower the NME; these negative contributions, once again, become more noticeable with increasing V$_{0pp}$.}

The GT transitions relevant for the $2\nu\beta\beta$ decay of $^{48}$Ca have been studied experimentally in Ref. \cite{Yako2009}. The GT$^-$ and GT$^+$ strength distributions in $^{48}$Sc have been measured by the $^{48}\text{Ca}(p,n)$ and $^{48}\text{Ti}(n,p)$ reactions, respectively. The integrated GT strengths up to an excitation energy of 30 MeV in $^{48}$Sc obtained from $(p,n)$ and $(n,p)$ spectra amount $\text{B}(\text{GT}^-)=15.3\pm2.2$ and $\text{B}(\text{GT}^+)=2.8\pm0.3$.
The REDF-QRPA calculations with DD-ME2 interaction for the corresponding transitions
result in $\text{B}(\text{GT}^-)= 23.47$ and  $\text{B}(\text{GT}^+)=3.48$, thus the experiment provides 65$\%$ of
GT$^-$ strength and 80$\%$ of the GT$^+$ strength obtained in model calculations.
Clearly, further experimental studies of GT transitions are needed to provide more 
transition strength that is relevant for double beta decays. The missing strength 
in measured GT spectra has been confirmed in recent REDF-QRPA calculations for other 
nuclei in Ref. \cite{Vale2020}, and studies going beyond the RPA level including
couplings between single nucleon and collective nuclear vibrations could not resolve the discrepancy between the theoretical and experimental GT 
strengths \cite{Niu2014PVC}.

\begin{figure}[h!]
\includegraphics[scale=0.45]{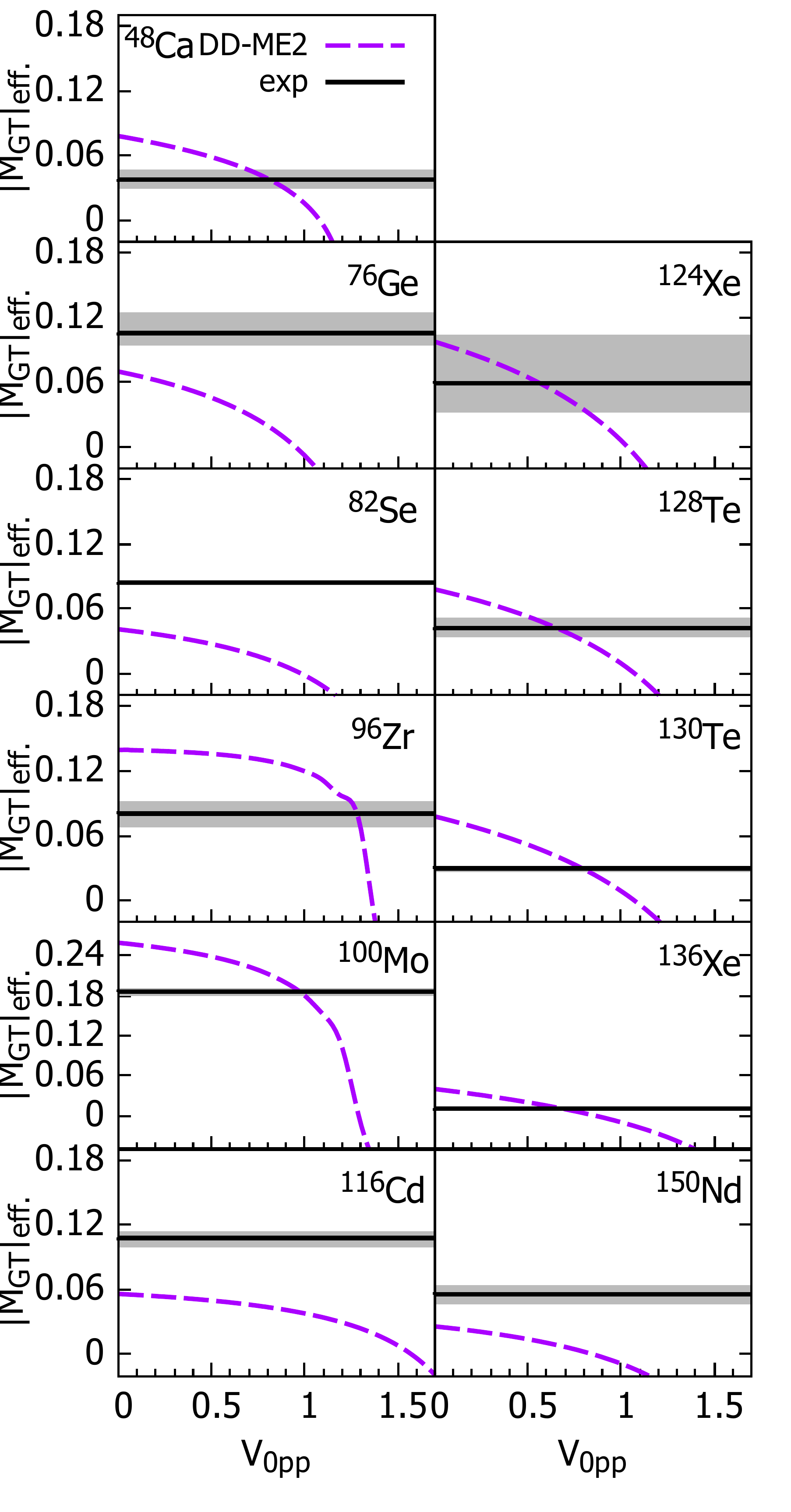}
\caption{The dependence of the NMEs for $2\nu\beta\beta$ decay on the isoscalar pairing 
strength $V_{0pp}$ for $48<A<150$ nuclei that decay through the $2\nu\beta\beta$ channel, in 
comparison to the result obtained from the experimental data \cite{Barabash20152}.
\label{fig6}}
\end{figure}

In the following, the NMEs are investigated for the set of nuclides usually considered in
$2\nu\beta\beta$ decay studies because there are experimental data available: $^{76}$Ge, $^{82}$Se, $^{96}$Zr, $^{100}$Mo, $^{116}$Cd, $^{128}$Te, ${}^{130}$Te, ${}^{136}$Xe, ${}^{150}$Nd, and most recently ${}^{124}$Xe. 
Figure \ref{fig6} shows the
NMEs for the set of 11 nuclides listed above, given as a function of the isoscalar pairing 
strength $V_{0pp}$.
The results extracted from the experimental data on $2\nu\beta\beta$ decay half-lives
are shown for comparison \cite{Barabash2015}.
We conclude that the dependence of the nuclear matrix elements on $V_{0pp}$ is qualitatively similar for all nuclei considered, with the values of the GT based NMEs decreasing and their
slopes increasing, with increasing absolute values of $V_{0pp}$.
However, there is a significant variation in the values of $V_{0pp}$ needed to reproduce the NMEs based on experimental data.
A similar issue has been observed in $\beta$-decay studies \cite{PhysRevC.93.025805},
indicating that some mass dependence is neccessary for the optimization of the isoscalar
pairing channel in the residual interaction of the REDF-QRPA, as we have also discussed in Sec. \ref{optimizeV0}.

In order to assess the information about the model dependence of the $2\nu\beta\beta$ decay
matrix elements, we conducted a detailed comparison of our results with a selection of
previous studies.
A direct comparison of the results of the present study with those from previous
calculations can be difficult both due to different theoretical frameworks and 
parametrisations used. 

%

In Ref. \cite{Pirinen2015}, the pn-QRPA matrix elements are given for several $A>100$ nuclides, using a value of the $T=0$ pairing strength $g_{pp}$ = 0.7. A comparison with the REDF-QRPA results with $V_{0pp}=$ 0 is given in Table \ref{table2}. We see that the upper limit of NMEs in the present study remains lower than those reported by other pn-QRPA calculations.
\begin{table}
\caption{The NMEs for 2$\nu\beta\beta$ decay based on the REDF-QRPA (DD-ME2 interaction, $V_{0pp}=$ 0) and the NMEs based on the pn-QRPA from Ref. \cite{Pirinen2015}.}
\label{table2}
\begin{ruledtabular}
\begin{tabular}{ ccc }
 & REDF-QRPA & pn-QRPA \cite{Pirinen2015}  \\
 & (DD-ME2) &  \\
\hline
$^{100}$Mo & 0.259 & 0.6560\\
$^{116}$Cd & 0.056 & 0.2169 \\
$^{128}$Te & 0.078 & 0.1041 \\
$^{130}$Te & 0.079 & 0.1066 \\
\end{tabular}
\end{ruledtabular}
\end{table}

Further comparison can be made with the recent pn-QRPA results of {\v S}imkovic et al.~\cite{Simkovic20182}, which is done in Table \ref{tableS}. We note that the values from Ref. \cite{Simkovic20182}, except for $^{48}$Ca and $^{82}$Se, are in excellent agreement with our results, even though the latter were obtained with vanishing isoscalar pairing. We also note that the NMEs given in Ref. \cite{Simkovic20182}  are effective matrix elements obtained with a quenched value of $g_A$ = 0.904. The renormalisation factor was split according to isospin channel into $g_{pp}^{T=1}$ and $g_{pp}^{T=0}$; the values of both parameters for each respective nucleus are given in Tab. \ref{tableS}.

\begin{table}
\caption{The NMEs for 2$\nu\beta\beta$ decay based on the REDF-QRPA (DD-ME2 interaction, $V_{0pp}=$ 0) and the NMEs based on the pn-QRPA from Ref. \cite{Simkovic20182} with the pairing strength parameters $g_{pp}^{T=1}$ and $g_{pp}^{T=0}$ given in the last two columns.}
\label{tableS}
\begin{ruledtabular}
\begin{tabular}{ ccccc }
 & REDF-QRPA & pn-QRPA \cite{Simkovic20182} & g$_{pp}^{T=1}$\cite{Simkovic20182} & g$_{pp}^{T=0}$\cite{Simkovic20182} \\
 & (DD-ME2) &  & & \\
\hline
$^{48}$Ca & 0.078 & 0.019 & 1.028 & 0.745 \\
$^{76}$Ge & 0.070 & 0.077 & 1.021 & 0.733 \\
$^{82}$Se & 0.041 & 0.071 & 1.016 & 0.737 \\
$^{96}$Zr & 0.140 &  0.162 & 0.961 & 0.739 \\
$^{100}$Mo & 0.259 & 0.306 & 0.985 & 0.799 \\
$^{116}$Cd & 0.056 & 0.059 & 0.892 & 0.877 \\
$^{128}$Te & 0.078 & 0.076 & 0.965 & 0.741\\
$^{130}$Te & 0.079 & 0.065 & 0.963 & 0.737\\
\end{tabular}
\end{ruledtabular}
\end{table}

Next, in Table \ref{table3} we compare the NMEs using the REDF-QRPA (DD-ME2 interaction, $V_{0pp}=$ 0 MeV) with those of recent Interacting Boson Model (IBM) calculations \cite{Barea2015,Iachello2015}. The IBM results have been calculated within a closure approximation, and to allow for a direct comparison, the values we report for the NMEs  have been obtained in the closure approximation as well. To obtain NMEs comparable to our own from the NMEs quoted by Barea and Iachello, which do not include the energy denominator, we divide the NMEs with an average energy denominator as tabulated in \cite{Kotila2012}.
The resulting NMEs based on the IBM are generally larger, but comparable to the REDF-QRPA results. The inclusion of isoscalar pairing in the REDF-QRPA residual interaction would further increase this difference. This result suggests an advantage of our calculations over the IBM, as the experimental values are even lower for most nuclei considered, and, for most nuclei, are within the reach for the REDF-QRPA with a suitable choice of $V_{0pp}$. 

Further comparison is made with the Interacting Shell Model (ISM) \cite{Coraggio2019}, as shown in Table \ref{table4}. {Since the ISM contains isoscalar pairing as a significant effect, we compare the shell model results to those of the REDF-QRPA calculated at the optimal pairing strength $V_{0pp}$ as given in Table \ref{tableV0pp}.
The REDF-QRPA results are considerably smaller than those of the ISM.}
%
\begin{table}
\caption{The NMEs for $2\nu\beta\beta$ decay based on the REDF-QRPA (DD-ME2 interaction, $V_{0pp}=$ 0), calculated in the closure approximation, and the results of the Interacting Boson Model (IBM) \cite{Iachello2015}.\label{table3}}
\begin{ruledtabular}
\begin{tabular}{ ccc }
 & REDF-QRPA & IBM \cite{Iachello2015} \\
 & (DD-ME2) &  \\
\hline 
$^{48}$Ca & 0.121 & 0.213 \\
$^{76}$Ge & 0.110 & 0.471 \\
$^{82}$Se & 0.071 & 0.356  \\
$^{96}$Zr & 0.125 & 0.208  \\
$^{100}$Mo & 0.210 & 0.272  \\
$^{116}$Cd & 0.056 & 0.197  \\
$^{128}$Te & 0.079 & 0.308  \\
\end{tabular}
\end{ruledtabular}
\end{table}

\begin{table}
\caption{The NMEs for $2\nu\beta\beta$ decay based on the REDF-QRPA (DD-ME2 interaction, optimal $V_{0pp}$), and the results of the Interacting Shell Model (ISM) (Ref. \cite{Coraggio2019}  unless otherwise noted). \label{table4}}
\begin{ruledtabular}
\begin{tabular}{ ccc }
 & REDF-QRPA & ISM (\cite{Coraggio2019}) \\
 & (DD-ME2) & \\
\hline 
$^{48}$Ca & 0.019 & 0.026 \\
$^{76}$Ge & 0.001 & 0.104 \\
$^{82}$Se &  0.001 & 0.109 \\
$^{124}$Xe\cite{CoelloPerez2019} & 0.023 & 0.028 --- 0.072 \\
$^{128}$Te \cite{Caurier2012} & 0.006 & 0.030 \\
$^{130}$Te & 0.002 & 0.061 \\
$^{136}$Xe \cite{Caurier2012} & 0.001 & 0.013 \\
\end{tabular}
\end{ruledtabular}
\end{table}

{More recent PN-QRPA results can be found in Ref. \cite{Deppisch2016}, but only for some of the nuclei considered in this work. The parameters for the calculation are determined using two procedures, fitting of the ft values to an isobaric triplet and multiplet. The comparison to the NMEs based on the REDF-QRPA with optimal $V_{0pp}$ is given in Table \ref{tablenew1}, showing reasonable agreement for $^{100}$Mo, while for
$^{116}$Cd and $^{128}$Te the REDF-QRPA provides smaller NMEs.}
\begin{table}
\caption{The NMEs for $2\nu\beta\beta$ decay based on the REDF-QRPA (DD-ME2 interaction, optimal $V_{0pp}$), and recent results of the proton-neutron QRPA (PN-QRPA) (Ref. \cite{Deppisch2016}. \label{tablenew1}}
\begin{ruledtabular}
\begin{tabular}{ cccc }
 & REDF-QRPA & PN-QRPA (\cite{Deppisch2016})& PN-QRPA (\cite{Deppisch2016}) \\
 & (DD-ME2) & triplet & multiplet \\
\hline 
$^{100}$Mo & 0.189 & 0.153 & 0.131 \\
$^{116}$Cd & 0.038 & 0.153 & 0.160 \\
$^{128}$Te & 0.006 & 0.069 & 0.095 \\
\end{tabular}
\end{ruledtabular}
\end{table}

Finally, we compare our calculations with the results of a recent effective theory (ET) treatment of double beta decay~\cite{CoelloPerez2018}. Effective theories concern the behaviour of nuclei at suitably low energies, which is described in terms of low-energy collective degrees of freedom, with any influence from high-energy physics being encoded into the low-energy constants which are fitted to available experimental data. The effective theory described in Ref. \cite{CoelloPerez2018} takes as its degrees of freedom nucleons and quadrupole phonon excitations of an even-even spherical core the nucleons are coupled to.
{The comparison of REDF-QRPA results using DD-ME2 interaction and optimal $V_{0pp}$, with those of the ET, is given in Table \ref{tableET}.} 
{With the exception of $^{100}$Mo, our results are systemically smaller, although they are comparable to the effective theory result in the case of $^{116}$Cd.}

\begin{table}
\caption{The NMEs for $2\nu\beta\beta$ decay based on the REDF-QRPA (DD-ME2 interaction, optimal $V_{0pp}$), and the results of the effective theory outlined in Ref \cite{CoelloPerez2018}. \label{tableET}}
\begin{ruledtabular}
\begin{tabular}{ ccc }
 & REDF-QRPA & ET (\cite{CoelloPerez2018}) \\
 & (DD-ME2) & \\
\hline 
$^{76}$Ge & 0.001 & 0.054 \\
$^{82}$Se &  0.001 & 0.097 \\
$^{100}$Mo & 0.189 & 0.111 \\
$^{116}$Cd & 0.038 & 0.085 \\
$^{128}$Te & 0.006 & 0.031 \\
$^{130}$Te & 0.002 & 0.021 \\

\end{tabular}
\end{ruledtabular}
\end{table}

\begin{figure}[h!]
\includegraphics[width=\linewidth]{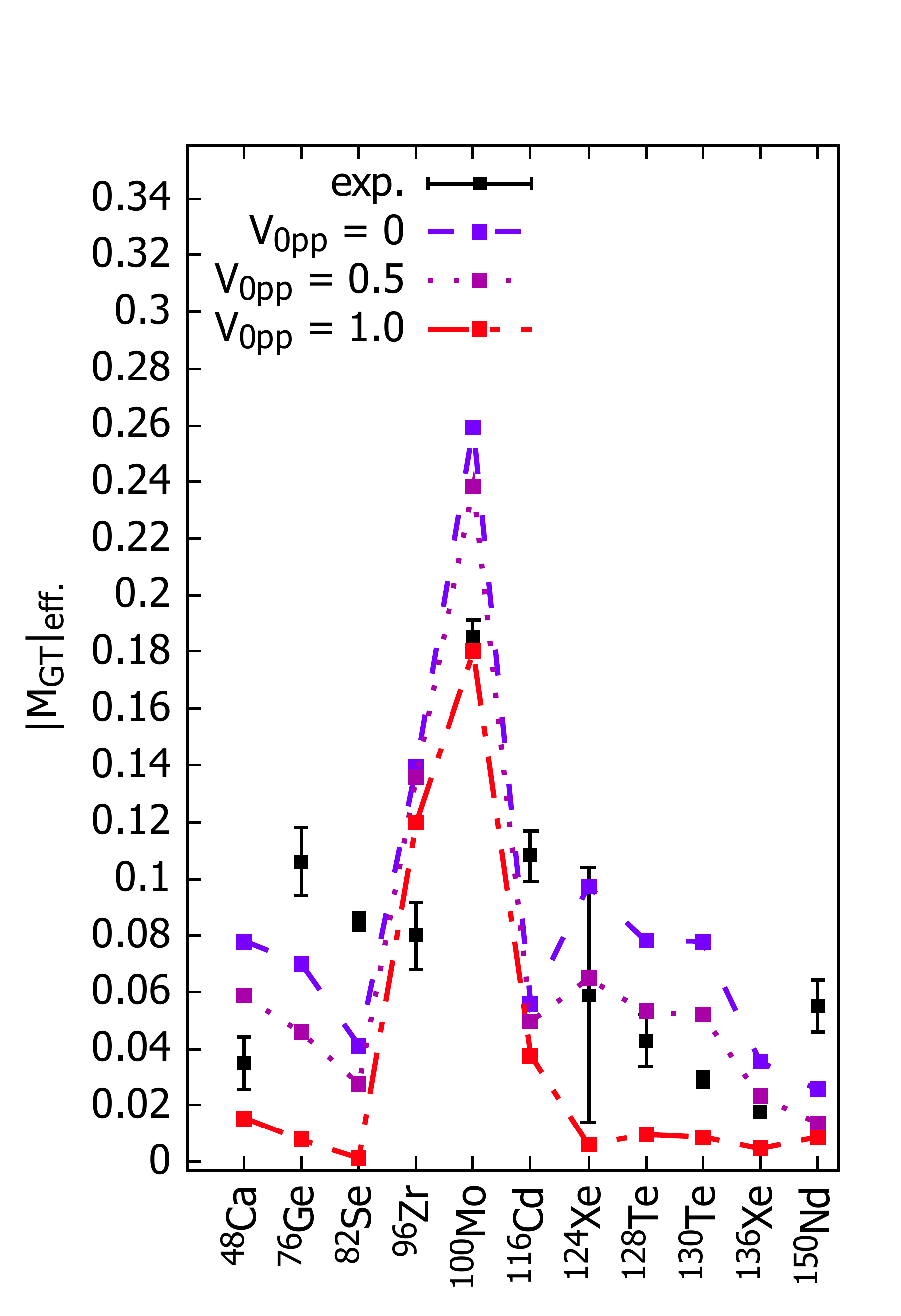}
\caption{The dependence of the NME for $2\nu\beta\beta$ decay on the isoscalar pairing strength $V_{0pp}$ for $A=48$--150 nuclides.\label{fig7}}
\end{figure}

One of the open questions in the description of double-beta decays is optimization of the
strength parameter of the isoscalar pairing in the residual QRPA interaction. Since this
interaction channel is difficult to constrain, and in most of the models cannot be 
determined based on the ground state properties, it is necessary to explore its
role in the NMEs for double-beta decays. In the present study, the experimental 
data on $\beta$-decay half lives are used to constrain the value of the isoscalar
pairing strength parameter $V_{0pp}$ (Sec. \ref{optimizeV0}), and in this way
the REDF-QRPA can provide predictions for $2\nu\beta\beta$ decay 
properties.
%
Fig. \ref{fig7} shows the NME for
$2\nu\beta\beta$ decay nuclides from $^{48}$Ca up to $^{150}$Nd, obtained using the REDF-QRPA with the DD-ME2 interaction, for the range of values for $V_{0pp}$ = 0--1. The experimental data adopted
 from Ref. \cite{Barabash20152} are also shown for comparison. One can observe
 systematic decrease of the NMEs with increasing the value of $V_{0pp}$. {For several
nuclei, the experimentally determined NMEs (with their error bars) are within the range of calculated NMEs when considering the full range of $V_{0pp}$ values from 0 to 1.
The exceptions are $^{76}$Ge, $^{82}$Se, $^{96}$Se, $^{116}$Cd, and $^{150}$Nd where the calculated NMEs are
 smaller (also the experimental errors are rather small), and $^{96}$Zr where the calculated NMEs are larger. In the case of $^{116}$Cd and $^{96}$Zr the differences between the experimental and calculated NMEs within given range of $V_{0pp}$ are very small.}
 
In Table \ref{table5} we give the NMEs for the DD-ME2 interaction,
using the optimal value for the isoscalar pairing strength $V_{0pp}$, in comparison 
to the values obtained from the experimental data on $2\nu\beta\beta$ decay \cite{Barabash20152}. The inclusion of $T=0$ pairing further reduces the
NMEs, thus allowing the agreement with the experimental data where previously the calculated NMEs have been too large. However, in {many} cases the isoscalar pairing reduced the NMEs such that the final
result is below the experimental range. 

\begin{table}
\caption{The NMEs for $2\nu\beta\beta$ decay obtained with the DD-ME2 interaction and optimal values of $V_{0pp}$ given in Tab. \ref{tableV0pp}, and NMEs obtained from the experimental data.\label{table5}}
\begin{ruledtabular}
\begin{tabular}{ ccc }
&  REDF-QRPA & exp. \cite{Barabash2020} \\
& (DD-ME2) & \\
\hline 
$^{48}$Ca & 		0.019$^{+	0.014	}_{-	0.019	}$ &  0.035 $\pm$ 0.003 \\
$^{76}$Ge & 		0.001$^{+	0.024	}_{-	0.001	}$ &  0.106 $\pm$ 0.004 \\
$^{82}$Se & 		0.001$^{+	0.011	}_{-	0.001	}$ &  0.085 $\pm$ 0.001 \\
$^{96}$Zr & 		0.121$^{+	0.006	}_{-	0.010	}$ &  0.080 $\pm$ 0.004 \\
$^{100}$Mo &		0.189$^{+	0.020	}_{-	0.033	}$ &  0.185 $\pm$ 0.002 \\
$^{116}$Cd &		0.038$^{+	0.004	}_{-	0.004	}$ &  0.108	 $\pm$ 0.003 \\
$^{124}$Xe &		0.023$^{+	0.021	}_{-	0.023	}$ &  0.059 $\pm$ 0.015 \\
$^{128}$Te &		0.006$^{+	0.011	}_{-	0.006	}$ &  0.043 $\pm$ 0.003 \\
$^{130}$Te &		0.002$^{+	0.010	}_{-	0.002	}$ &  0.0293 $\pm$ 0.0009 \\
$^{136}$Xe &		0.001$^{+	0.004	}_{-	0.001	}$ &  0.0177 $\pm$ 0.0002 \\
$^{150}$Nd &		0.013$^{+	0.007	}_{-	0.005	}$ &  0.055 $\pm$ 0.003 \\
\end{tabular}
\end{ruledtabular}
\end{table}

\begin{figure}[h!]
\includegraphics[width=\linewidth]{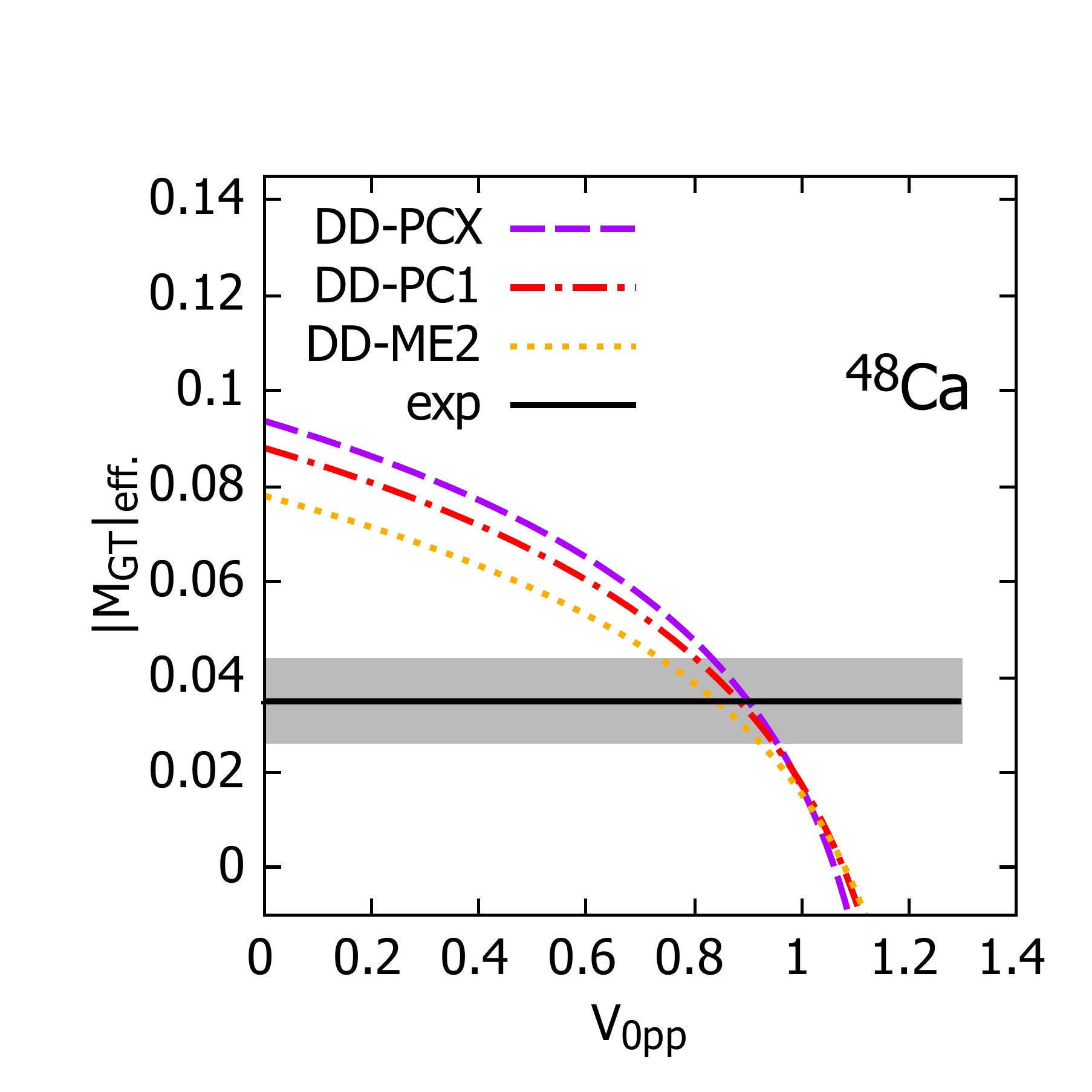}
\caption{Comparison of the NMEs for $^{48}$Ca using the DD-PC1 and DD-PCX interactions and the value obtained from experimental data \cite{Barabash2020}. \label{fig8}}
\end{figure}

\begin{figure}[h!]
\includegraphics[width=\linewidth]{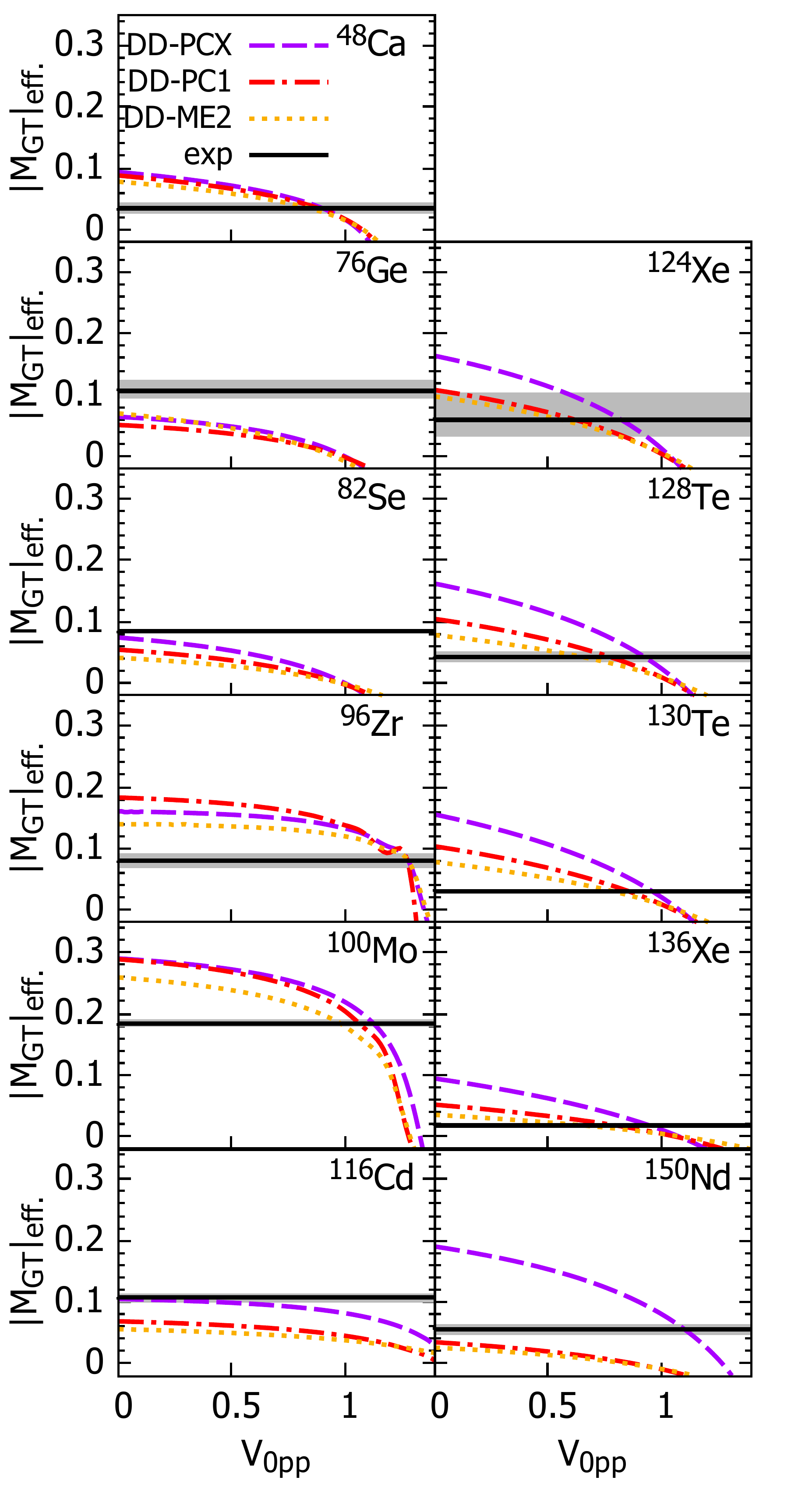}
\caption{The dependence of the NMEs on the isoscalar pairing strength $V_{0pp}$ for the $2\nu\beta\beta$ decay for the set of nuclides in the mass range $^{76}$Ge--$^{150}$Nd, {for all relativistic interactions employed in this work} shown in comparison to the values obtained from the experimental data \cite{Barabash2020}. \label{fig9}}
\end{figure}

In order to assess the information about the systematic model dependence of the NMEs for $2\nu\beta\beta$ decay  in the REDF framework, we extend our calculations also to the relativistic point coupling interactions (see Sec. \ref{relativistic}). 
In Fig. \ref{fig8} the NMEs are shown for $^{48}$Ca, obtained using point coupling interactions DD-PC1 and DD-PCX, those with meson exchange
interaction DD-ME2, and the NME obtained from the experimental data on $2\nu\beta\beta$ decay \cite{Barabash20152}. The figure shows the dependence of the NMEs on the $T=0$ pairing strength $V_{0pp}$. 
The results for the two point coupling interactions display some variations of the NMEs, though not considerable; the NMEs for the DD-PC1 interaction are somewhat lower, e.g. at $V_{0pp}=0$ the difference in the NMEs for DD-PC1 and DD-PCX interactions is less than $\approx0.01$. The NMEs for DD-ME interaction show
qualitatively the same dependence on $V_{0pp}$.
Similar analysis of the sensitivity of the NMEs on the effective interaction employed is performed for other nuclei of interest, from $^{76}$Ge toward $^{128}$Te, as shown in Fig. \ref{fig9}. The NMEs for DD-PC1 interaction appear systematically smaller than those of DD-PCX interaction, and their differences provide the insight into the model uncertainties when using different point coupling interactions. In the case of DD-ME2 interaction, the NMEs are smaller than those of DD-PC1 and DD-PCX interactions, except for $^{76}$Ge.The differences depend on the specific nucleus under consideration. For example, for $^{48}$Ca the results for the two interactions
almost coincide, while for $^{150}$Nd considerable differences in the NMEs are obtained.

\begin{table}
\caption{The same as Tab. \ref{table5} but for DD-PC1 and DD-PCX interactions and the corresponding
optimal values of the isoscalar pairing strength $V_{0pp}$ given in Tab. \ref{tableV0pp}.\label{tablenew11}}
\begin{ruledtabular}
\begin{tabular}{ cccc }
& REDF-QRPA & REDF-QRPA & exp. \cite{Barabash20152} \\
& (DD-PC1) & (DD-PCX) & \\
\hline 
$^{48}$Ca & 0.008$^{+	0.019		}_{-	0.008	}$ & 0.024$^{+	0.016	}_{-	0.021	}$ & 0.035 $\pm$ 0.003 \\
$^{76}$Ge & 0.003$^{+	0.021		}_{-	0.003	}$ & 0.009$^{+	0.014	}_{-	0.009	}$ & 0.106 $\pm$ 0.004 \\
$^{82}$Se & 0.009$^{+	0.019		}_{-	0.009	}$ & 0.005$^{+	0.014	}_{-	0.005	}$ & 0.085 $\pm$ 0.001 \\
$^{96}$Zr & 0.148$^{+	0.016		}_{-	0.027	}$ & 0.160$^{+	0.007	}_{-	0.011	}$ & 0.080 $\pm$ 0.004 \\
$^{100}$Mo &0.188$^{+	0.028		}_{-	0.049	}$ & 0.233$^{+	0.016	}_{-	0.026	}$ & 0.185 $\pm$ 0.002 \\
$^{116}$Cd &0.042$^{+	0.006		}_{-	0.008	}$ & 0.087$^{+	0.005	}_{-	0.006	}$ & 0.108	 $\pm$ 0.003 \\
$^{124}$Xe &0.034$^{+	0.036		}_{-	0.019	}$ & 0.034$^{+	0.084	}_{-	0.034	}$ & 0.059 $\pm$ 0.015 \\
$^{128}$Te &0.009$^{+	0.015		}_{-	0.005	}$ & 0.012$^{+	0.015	}_{-	0.012	}$ & 0.043 $\pm$ 0.003 \\
$^{130}$Te &0.014$^{+	0.015		}_{-	0.013	}$ & 0.006$^{+	0.018	}_{-	0.006	}$ & 0.0293 $\pm$ 0.0009 \\
$^{136}$Xe &0.007$^{+	0.006		}_{-	0.005	}$ & 0.003$^{+	0.008	}_{-	0.003	}$ & 0.0177 $\pm$ 0.0002 \\
$^{150}$Nd &0.023$^{+	0.010		}_{-	0.008	}$ & 0.066$^{+	0.019	}_{-	0.022	}$ & 0.055 $\pm$ 0.003 \\
\end{tabular}
\end{ruledtabular}
\end{table}

\begin{figure}[h!]
\includegraphics[width=\linewidth]{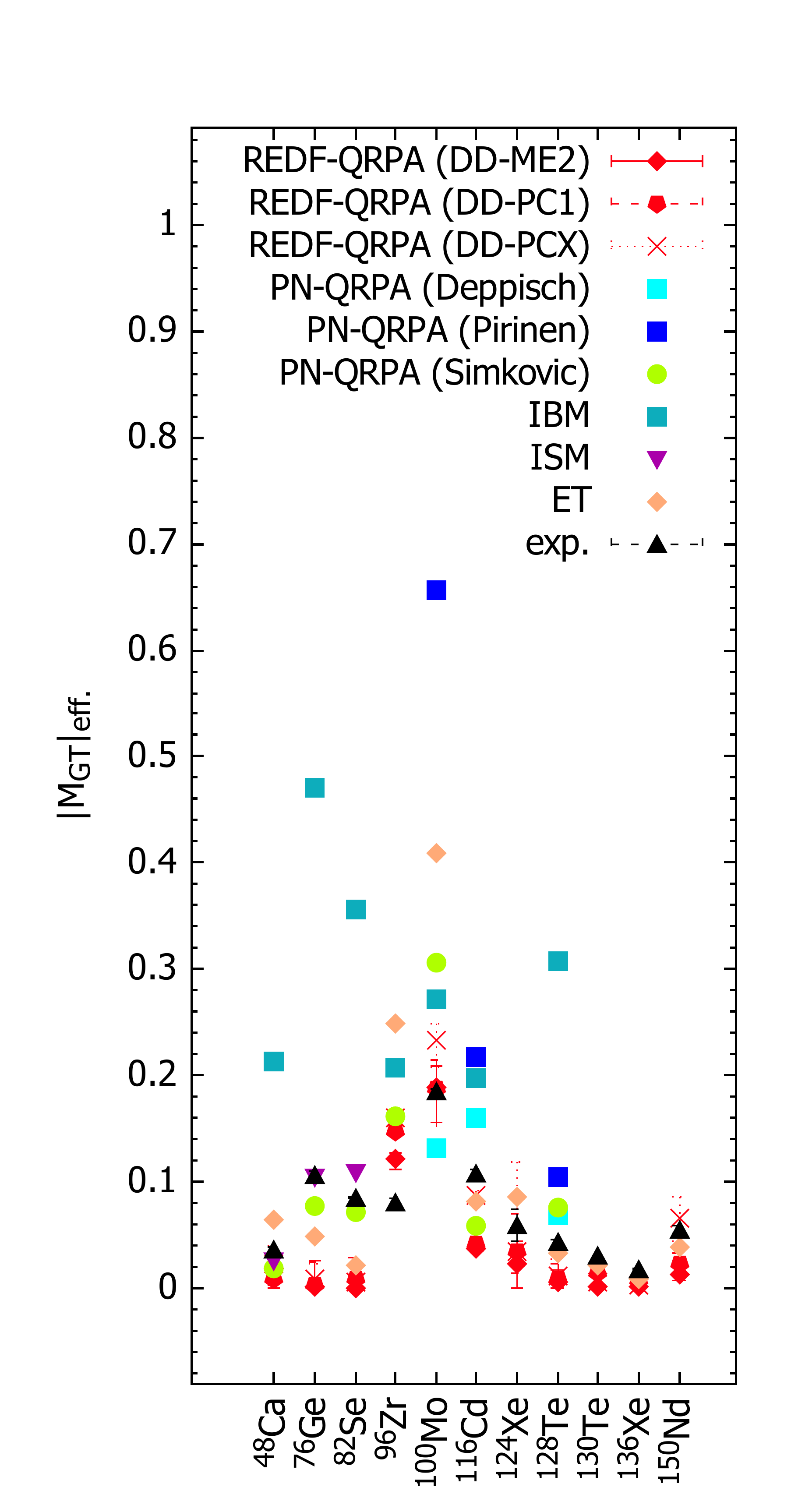}
\caption{Summary of the $2\nu\beta\beta$ decay NMEs from the REDF-QRPA with DD-ME2, DD-PC1 and DD-PCX interactions with optimal $T=0$ pairing together with respective $1\sigma$ uncertainties, compared to the calculations based on the pn-QRPA by Deppisch and Suhonen ~\cite{Deppisch2016}, Pirinen~\cite{Pirinen2015}, and  {\v S}imkovic~\cite{Simkovic20182}, Interacting Boson Model (IBM)~\cite{Iachello2015}, Interacting Shell Model (ISM) ~\cite{Coraggio2019}, and the effective theory (ET) outlined in \cite{CoelloPerez2018}. The NMEs from the experimental data~\cite{Barabash20152} are also shown. \label{fig10}}
\end{figure}

The optimal values for the $T=0$ pairing strength, obtained for the REDF-QRPA with DD-PC1 and DD-PCX interactions (Table \ref{tableV0pp}), can now be employed in description of $2\nu\beta\beta$ decay NMEs.  The calculated NMEs are summarized in Table \ref{tablenew11} for DD-PC1 and DD-PCX interactions, respectively, in comparison to the experimental values. Although there are some variations in the NMEs when compared to the experimental data, an overall reasonable 
qualitative agreeement is obtained.
This is illustrated in Fig. \ref{fig10}, where we summarize the results of the present study, including the NMEs  for $2\nu\beta\beta$ decay obtained using the REDF-QRPA with DD-ME2, DD-PC1, and DD-PCX interactions with the corresponding optimal values of isoscalar pairing strength parameters. For comparison, the NMEs from previous studies are shown, including pn-QRPA implementations by Suhonen \cite{Suhonen2005}, Pirinen \cite{Pirinen2015}, and {\v S}imkovic \cite{Simkovic20182}, Interacting Boson Model (IBM) \cite{Iachello2015} and the Interacting Shell Model (ISM) \cite{Coraggio2019}, as well as the experimental result \cite{Barabash20152}. 
Clearly, the results of the REDF-QRPA are of comparable quality than those from previous studies, though for most of nuclei somewhat lower than those determined
from the experimental data. In addition, rather than providing just a single NME value, our study provides the insight into systematic uncertainties due to variations in
the formulation of the REDF and parametrizations used. Additional uncertainty is accounted for due to the isoscalar pairing strength interaction that is constrained 
by $\beta$-decay half-lives.

{The nuclear matrix elements calculated in this work appear rather small compared to 
previous studies. As already discussed, the isoscalar and isovector pairing strengths
are very similar, indicating that the spin-isospin SU(4) symmetry is softly broken.
Weak breaking of the SU(4) symmetry could partly explain small NMEs obtained 
within REDF-QRPA approach, but even for $V_{0pp}=0$ the the NMEs appear smaller
than in other approaches. Another issue is that our treatment did not include
relevant effects going beyond the QRPA, as mentioned in the Sec.\ref{Intro}.}

\begin{figure}[h!]
\includegraphics[width=\linewidth]{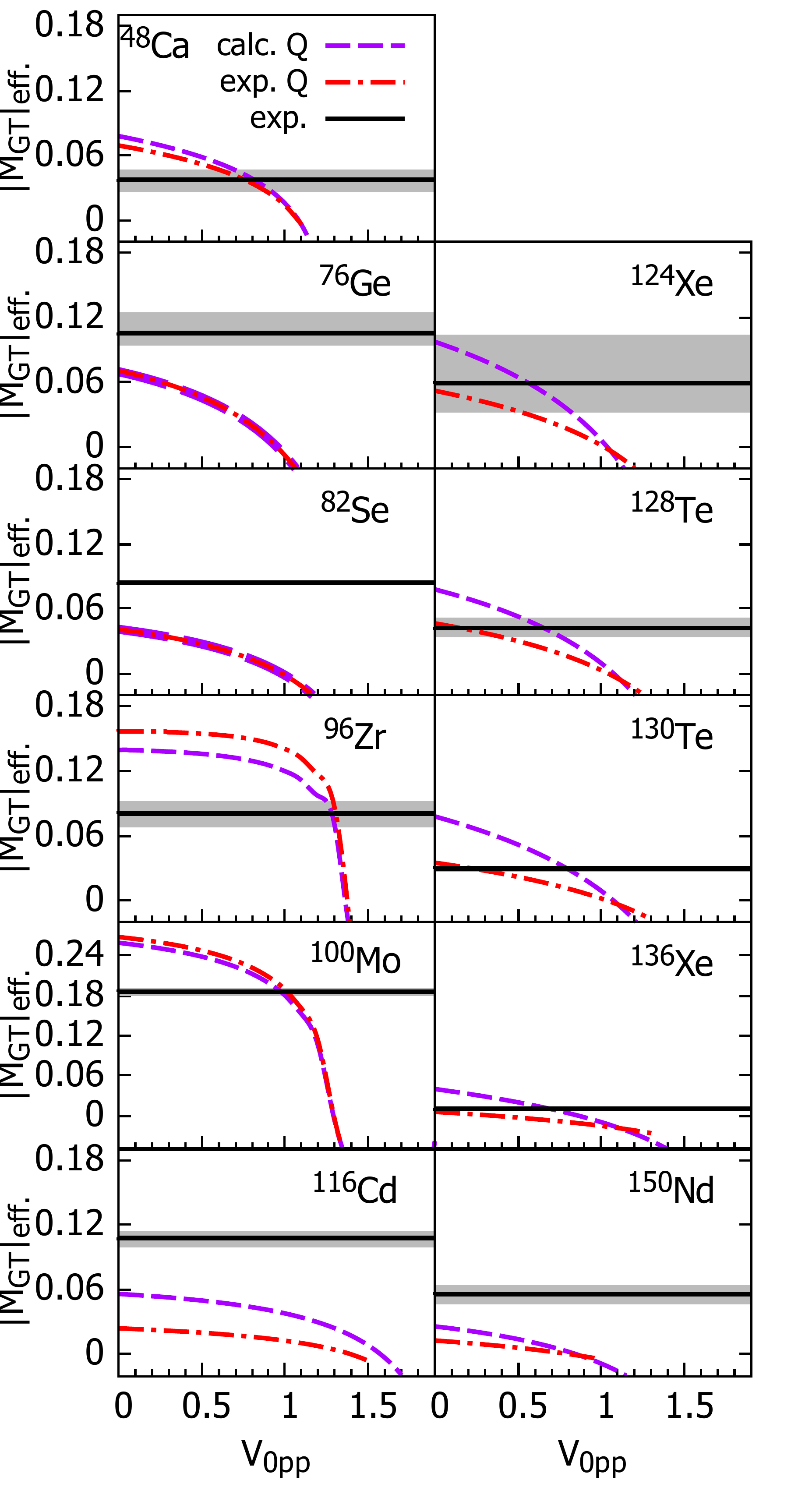}
\caption{{The NMEs for the DD-ME2 interaction, calculated with the Q-values derived self-consistently 
in the RH-BCS framework, shown next to the NMEs for the same interaction, but with the Q-values based on 
experimental data \cite{Wang2021}.} \label{fign1}}
\end{figure}

{In order to preform an additional sensitivity check of our calculations we have also preformed calculations in which Q values derived from the experimenal masses have been used. Figure \ref{fign1} shows the NMEs
calculated using DD-ME2 interaction, both with experimental and calculated Q values.
One can observe that the NMEs results are nearly the same for $^{48}$Ca, $^{76}$Ge, and $^{82}$Se, while for other nuclei the same trend with $V_{0pp}$ is obtained, but some differences can be
observed. In several cases the NMEs become smaller when using experimental Q values instead of calculated
ones. Thus, the choice of Q value does not provide a solution to the problem of small NMEs values, except for $^{96}$Zr.}
{Whether the 2$\nu\beta\beta$ NMEs could be increased by an appropriate treatment
of the effects we could not include in the current work is an interesting perspective for the further research.
For example, in Ref. \cite{Rodriguez2017} the structure of $^{76}$Ge and $^{76}$Se, has been studied with Gogny functionals,
indicating that the comparison with the experimental data could only be obtained when triaxial shapes have been included, and these are precisely the nuclei for which we obtain the NMEs that diverge the most from experimental results.
Also in the present study the systematic uncertainty due to implementation of three different relativistic interactions is considerably larger than the experimental uncertainty. Therefore, further investigations and improvements of the theory frameworks are necessary in order to reduce the theoretical uncertainty 
obtained from our, but also from other studies.}
 
\section{Conclusion}

In this  work a theory framework is established for the study of $2\nu\beta\beta$ decay nuclear matrix elements based on the relativistic nuclear energy functional. Model calculations include two different formulations of the effective interactions, density-dependent meson-exchange and point coupling interactions, and three parameterizations (DD-ME2, DD-PC1, DD-PCX) have been employed in order to assess the information on the systematic uncertainties on the NMEs in the relativistic framework. The ground states of nuclei involved in the decay
are calculated within the relativistic Hartree-BCS model, while 
nuclear transitions in the $2\nu\beta\beta$ decay are described using the REDF-QRPA.
In addition to the isovector pairing correlations taken into account in the ground state 
calculations within the RH-BCS model, the isoscalar paring channel has also been included in 
the residual REDF-QRPA interaction, and its optimal values of the strength parameter $V_{0pp}$ are
constrained by the experimental data on $\beta$-decay half lives, in order to allow predictions in 
double beta decay studies.

We have calculated the NMEs for a set of nuclei
that undergo $2\nu\beta\beta$ decay, $^{48}$Ca, $^{76}$Ge, $^{82}$Se, $^{96}$Zr, $^{100}$Mo, $^{116}$Cd, $^{128}$Te, ${}^{130}$Te, ${}^{124}$Xe, ${}^{136}$Xe, and ${}^{150}$Nd. {The
dependence of the NMEs on the isoscalar paring strength $V_{0pp}$ has been investigated
at the limit $V_{0pp}=0$ and using optimal $V_{0pp}$ values.}
%
The NMEs from the REDF-QRPA provide an improvement
over the interacting boson model, being closer to the values of the NMEs obtained from the experimental data. 
{However, when compared to the non-relativistic pn-QRPA, interacting shell model, or effective theory, the NMEs from the present study for most of studied
nuclei, with a few exceptions, are rather small. While for some studied nuclei 
the NMEs are already below experimental ones in the $V_{0pp}=0$ limit, 
for several nuclei at this limit the NMEs are above experimental values, 
but when introducing the optimal value of $V_{0pp}$, the NMEs become rather low.
It has been shown that different
treatment of the Q value calculation for most of studied nuclei could not increase 
the NMEs, and further studies of additional effects are required to resolve 
this question.}
Rather than providing just a single NME value like most of previous studies, our work 
provides the insight into systematic uncertainties due to different formulations of the REDF
and parametrizations used. Additional uncertainty is accounted for due to the isoscalar pairing strength
interaction that is constrained by $\beta$-decay half-lives.

This work provides an important benchmark for the future applications
of the relativistic framework in studies of neutrinoless double-beta decay.
However, we note that the present study represents our first study
of double-beta decays in the relativistic framework, and some effects have not been
considered, e.g., nuclear deformation, symmetry restoration, {configuration mixing}, etc. Future improvements of
the REDF based theory framework, in particular the on-going development of the 
deformed REDF-QRPA for the functionals used in this work, will allow
additional improvements in modeling nuclear double-beta decays. 

\section{Acknowledgements}
We thank Deni Vale for support regarding 
the REDF-QRPA and useful discussions, and Jenni Kotila for discussion on the phase space factors for
double beta decays.
This work is supported by the QuantiXLie Centre of Excellence, a project co-financed by the Croatian Government and European Union through the European Regional Development Fund, the Competitiveness and Cohesion Operational Programme (KK.01.1.1.01.0004).

\cleardoublepage
\bibliography{articleRevTex.bbl}

\end{document}